# Evaluating LLMs for Career Guidance: Comparative Analysis of Computing Competency Recommendations Across Ten African Countries

Precious Eze, Stephanie Lunn, Bruk Berhane School of Universal Computing, Construction and Engineering Education (SUCCEED), College of Engineering and Computing, Florida International University, 10555 West Flagler Street, Miami, FL 33174, U.S.A.

E-mail: peze001@fiu.edu, sjlunn@fiu.edu, bberhane@fiu.edu

**Abstract**

Employers increasingly expect graduates to utilize large language models (LLMs) in the workplace, yet the competencies needed for computing roles across Africa remain unclear given varying national contexts. This study examined how six LLMs, namely ChatGPT 4, DeepSeek, Gemini, Claude 3.5, Llama 3, and Mistral AI, describe entry-level computing career expectations across ten African countries. Using the Computing Curricula 2020 framework and drawing on Digital Colonialism Theory and Ubuntu Philosophy, content analysis of 60 LLM responses to standardized prompts reveals consistent coverage of technical competencies such as cloud computing and programming, but notable differences in non-technical competencies, particularly ethics and responsible AI use. Models vary considerably in recognizing country-specific factors, including local technology ecosystems, language requirements, and national policies averaging only 35.4% contextual awareness overall. Open-source models demonstrated stronger contextual awareness and better balance between technical and professional skills, with Llama (4.47/5) and DeepSeek (4.25/5) outperforming proprietary alternatives ChatGPT-4 (3.90/5) and Claude (3.46/5). However, Mistral's poor contextual performance (0.00/4) despite being open-source indicates that development philosophy alone does not guarantee contextual responsiveness. This first comprehensive comparison of LLM career guidance for African computing students uncovers entrenched infrastructure assumptions and Western-centric biases that create gaps between technical recommendations and local realities. The findings challenge assumptions about AI tool quality in resource-constrained settings and underscore the need for decolonial approaches to AI in education, emphasizing contextual relevance and hybrid human-AI guidance models.

**Keywords:** Large Language Models; Career Guidance; Computing Education; African Higher Education; Competency Frameworks; Educational Technology; Digital Colonialism; Ubuntu Philosophy; AI Contextual Adaptation

## 1. Introduction

Artificial intelligence (AI) and education have converged at a pivotal moment both globally and across Africa, creating substantial changes in how learning and career development unfold. Large Language Models (LLMs), a subset of generative AI systems that use machine learning to produce human-like responses from massive training datasets, are becoming more common, especially where students have limited access to professional academic guidance (Brachman et al., 2025). Students and recent graduates now turn to these conversational AI tools for help with coursework, career planning, job applications, and skill development (Akiba & Fraboni, 2023; Chang et al., 2024). This shift is especially pronounced in computing education, where competencies are defined as the integrated combination of skills, knowledge, and professional dispositions required for effective workplace performance. The *Computing Curricula 2020* framework, developed by the Association for Computing Machinery, the Institute of Electrical and Electronics Engineers, and other partners, provides a globally recognized standard for these competencies (Clear et al., 2020). These competencies now encompass both traditional technical abilities and emerging areas such as AI literacy, thus shaping employers' expectations for computing graduates in relation to these skills (Murire, 2024; World Economic Forum, 2023).

However, this technological shift is not without tension. Most LLMs are developed using training data that mirror Western educational norms, values, technological infrastructure, and labor market assumptions (Shahid et al., 2025). Birhane (2020) describes this dynamic as algorithmic colonization, wherein Western-developed AI tools are imported to Africa without adaptation, often producing outcomes that do not truly serve the local communities adopting these tools. Similarly, Mohamed et al. (2020) argue that AI systems risk perpetuating colonial patterns of power when they are deployed without attention to local context. These concerns carry direct implications for how African educators, institutions, and policymakers approach AI tools.

Recent evidence suggests AI adoption is growing across Africa. The 2025 Stanford AI Index notes that two-thirds of countries now offer or plan to offer AI-related courses within K-12 (primary and secondary) computer science education, twice as many as in 2019 (Maslej et al., 2025). Among the countries mentioned, Africa and Latin America were considered to be making the most progress, though it was acknowledged that access remains limited in many African countries due to basic infrastructure gaps such as electricity. Meanwhile, the African Union Executive Council endorsed the Continental AI Strategy during its 45th Ordinary Session in July 2024, underscoring Africa's commitment to an Africa-centric, development-focused approach to AI (African Union, 2024).

Computing education research supports these concerns regarding LLMs' reliability for contextually relevant guidance. Lunn et al. (2021) noted that securing and thriving in a computing role requires more than technical skills alone; it demands the development of a professional identity, gaining industry recognition, and learning to navigate the social and institutional relationships that define particular computing industries and ecosystems. This has far-reaching implications for Africa, where countries like Nigeria, Kenya, and South Africa have become key players in the continent's growing technology sector, attracting increased attention from global investors (African Union, 2024). African job market projections reveal that the continent will have the largest working-age population by 2040 (McKinsey Global Institute, 2023), and surveys indicate that students in African countries like Uganda extensively engage with AI tools for learning and job preparation (Ajalo et al., 2025). While the African Union has highlighted AI's role in refining workforce skills and driving its development (African Union, 2024; United Nations, 2024), there remains limited consensus on the exact mix of technical and non-technical competencies that computing graduates need to thrive. Moreover, the competencies required may vary with regard to specific computing roles. This uncertainty reinforces the need to critically assess how well LLM-generated advice aligns with Africa's distinct educational and workforce requirements.

Early studies have shown AI's potential in career planning for students entering the workforce (Westman et al., 2021), but existing scholarship on AI in career guidance has focused almost exclusively on Western contexts (Brandenburger & Janneck, 2023; Leung, 2022). Current empirical studies on computing education in Africa repeatedly point to the need for learning that fits individual student needs and bridges skill gaps in curricula (Hamouda et al., 2025). Ade-Ibijola and Okonkwo (2023) argued that despite increased adoption of AI-driven education in Africa, many graduates still struggle to transition into the workforce due to a misfit between skills acquired through formal education and employer demands. There is currently no comprehensive study comparing how different LLMs provide career guidance for computing students across African countries, yet the need for such comparison is clear given the wide differences in education systems, access to technology, and cultural contexts across the continent.

### 1.1 Research Objectives

Recognizing that our understanding of LLM-generated responses may be limited in specific contexts, this study explored how six LLMs respond to questions about entry-level skills and competencies for computing graduates across ten African countries. Central to this examination is the distinction between proprietary and open-source models. Proprietary LLMs typically require expensive subscriptions and function as closed systems that restrict adaptation for specific contexts (Bommasani et al., 2021), while open-source alternatives can be downloaded, hosted locally, and fine-tuned to institutional needs (Touvron et al., 2023). This distinction is relevant for resource-constrained African institutions needing models that are efficient, accessible, and locally adaptable (Tonja et al., 2024). As LLMs become more common in educational and professional environments, computing graduates are expected to use these tools effectively and understand the ethical problems and limitations that come with using them at work (Baidoo-Anu & Owusu Ansah, 2023; Brachman et al., 2025). A recent study looking at the perception of African students and graduates towards AI-driven career guidance found that participants were receptive to AI tools and considered them valuable in filling the gaps in traditional career guidance, with identical perception across genders regarding the efficacy of these tools (Eze et al., 2025). These findings point to an evolving skill set that is becoming crucial for African computing graduates to stay competitive in a rapidly changing job market.

To achieve our research objectives, ten countries in Africa were selected based on the Oxford Insights Government AI Readiness Index (2024), ensuring representation across diverse linguistic, cultural, and economic contexts. The countries examined included: Egypt, South Africa, Tunisia, Morocco, Nigeria, Senegal, Kenya, Benin, Ghana, and Zambia. Mauritius and Rwanda, while ranking high on the AI Readiness Index, were not included in the final sample. This omission resulted from an oversight during the initial country mapping phase, which was only identified after data collection and analysis had been completed. However, the ten countries retained in the study capture substantial diversity across Anglophone, Francophone, and Arabophone contexts, spanning a range of AI readiness levels from Egypt at 75th globally to Morocco at 124th, and collectively representing the educational and labor market conditions that this research seeks to examine.

Accordingly, the study addresses the following research questions (RQs):

**RQ1:** What are the similarities in how LLMs describe technical and non-technical computing competencies associated with LLM use across the selected countries?

**RQ2:** What are the differences in how LLMs describe technical and non-technical computing competencies associated with LLM use across ten selected countries?

In evaluating LLM performance, this study examines three dimensions: technical coverage refers to the breadth and depth of computing competencies addressed relative to the CC2020 framework (Clear et al., 2020); contextual awareness measures the incorporation of local industry references, language considerations, national policies, institutional factors; and overall relevance assesses the integration of technical coverage, contextual awareness, balance between technical and non-technical skills and the depth of LLM response. Consequently, the third research question is:

**RQ3:** How do different LLMs compare in their technical coverage, contextual awareness, and overall relevance of career guidance for these contexts?

The paper proceeds as follows: Section 2 reviews relevant literature on LLMs in education and African computing contexts; Section 3 presents our theoretical and analytical frameworks; Section 4 details the methodology; Section 5 reports findings across technical, contextual, and comparative dimensions; Section 6 discusses theoretical and practical implications; Section 7 addresses limitations and future directions; and Section 8 concludes.

## 2. Literature Review

The following review establishes the foundation for understanding how LLMs function as career guidance tools and their implications for African computing education. We look at how LLMs have developed in educational settings to show their growing influence on career preparation, examine the particular challenges and opportunities that define AI adoption across African computing education, and identify theoretical perspectives that help explain the tensions between globally-developed AI systems and local educational realities.

### 2.1 LLMs in Education and Career Guidance Contexts

AI in education has evolved rapidly, especially since transformer-based language models emerged in 2022. The attention mechanism introduced by Vaswani et al. (2017) allowed current AI models to process and generate large text corpora, leading to more advanced applications of AI tools in education. Five key areas where LLMs are being applied in education have been identified: tutoring, grading, content creation, language learning, and career guidance (Wang et al., 2024). Within computing education, these tools are reshaping fundamental pedagogical practices, requiring educators to navigate unprecedented challenges around academic integrity, assessment validity, and skill development (Prather et al., 2023).

Recent large-scale work has started documenting how students perceive and use these tools, though the evidence base remains geographically uneven. A survey of 23,218 higher education students across 109 countries found that students primarily use ChatGPT for brainstorming, summarizing texts, and finding research articles, viewing it as useful for simplifying complex information but less reliable for supporting classroom learning (Ravšelj et al., 2025). However, this sample drew heavily from European and North American institutions, leaving open the question of whether these patterns hold in African contexts where different educational traditions and infrastructure constraints shape how students engage with AI tools. A study of 453 students in the UK found that while over half had used AI tools for their degrees, barriers remain, including fear of plagiarism accusations and uncertainty about acceptable use (Thomson et al., 2024). Two-thirds believed ethical AI use would matter for their future careers, yet most were unaware of their institutions' AI policies, a gap likely to be wider in African institutions where AI governance frameworks are still emerging.

Generative AI is becoming more popular in career guidance. Recent work has demonstrated how the Gemini Pro API can create detailed career roadmaps based on student interests and aptitudes, providing specific advice about competitive exams, college choices, and career paths (Waikar et al., 2024). Similarly, generative AI has been argued to move vocational education away from traditional "one-size-fits-all" career guidance toward personalized approaches that adjust to individual learning styles and evolving job market needs (Duan & Wu, 2024). Focus groups with students and guidance staff in Finland identified key requirements for AI in career guidance, including attention to ethical issues like privacy, equality, and transparency (Westman et al., 2021). This study predates ChatGPT and the current generation of LLMs, reflecting expectations about AI capabilities rather than direct experience with tools currently in use, though the rapid evolution of AI means foundational considerations around ethics and personalization remain relevant even as specific tools change. What remains unclear is whether the collaborative human-AI models for career guidance are feasible in contexts where access to trained career counselors is already limited; evidence suggests that computing students often seek tech-specific career support from organizations and clubs rather than academic advisors (Lunn et al., 2022). Given this context, AI-driven career guidance has attracted considerable attention, leading researchers and educators to look at how AI tools like LLMs can improve learning and career development in different educational settings.

Educational institutions and researchers have started building specialized AI platforms that turn these capabilities into practical career support systems. ResumeAI, an AI-based platform, helps users tailor their résumés for specific job requirements by identifying skill gaps and matching content to employer expectations (Rahman et al., 2023). Similarly, an AI-powered platform built for Sri Lankan university students uses Graph Neural Networks for course recommendations and Recurrent Neural Networks for career guidance, reaching 81.88% accuracy in creating personalized career pathways (Dewasurendra et al., 2024). A study evaluating gamified AI use among grade 7 to 9 students found that combining interactive AI, role-play, and real-world data in a game-like system improved how students think about career development (Brandenburger & Janneck, 2023). Together, these studies show that AI-driven career guidance systems represent a shift from generic counseling to flexible, data-driven support that adjusts to individual student needs. Recent studies are beginning to challenge how we think about generative AI in career guidance (Zheng et al., 2023; Bhattacharjee et al., 2024; World Economic Forum, 2023). Rather than viewing them simply as tools for delivering information, these studies explore their potential to support personalized, context-aware guidance. AI models can work as partners throughout students' academic experience, providing tailored advice for course selection and professional development (Westman et al., 2021). LLMs have proven useful in academic advising, making information easier to understand for students from different backgrounds (Akiba & Fraboni, 2023). However, scholars argue that AI should not replace human advisers who better understand emotional, social, and contextual factors but rather work as a decision support tool (Leung, 2022; Sampson & Osborn, 2015).

While these studies highlight AI's potential, they tend to assume that LLM outputs are neutral conduits of information. A growing body of work challenges this assumption. Research mapping how biases emerge at each stage of the LLM life cycle, from training data collection through deployment, shows that traditional machine learning bias measures do not transfer directly to LLM-generated text (Lee et al., 2024). This framework identifies representation bias in training corpora, learning bias in model optimization, and deployment bias in how outputs are interpreted, each stage introducing distortions that compound in the final output. For career guidance, this means that even well-intentioned recommendations may reflect labor market assumptions and career trajectories drawn primarily from Western contexts where training data is concentrated.

A different concern relates not to bias in the advice itself, but to changes in how students engage with the advising process. An experimental study of 117 university students found that ChatGPT users produced better essay scores but showed no advantage in actual knowledge gain or transfer (Fan et al., 2025). More concerning, the analysis revealed "metacognitive laziness," reduced self-regulation, and fewer evaluation processes when using AI tools. If this pattern extends to career decision-making, students using LLM-based guidance may arrive at choices without developing the reflective capacity to evaluate whether those choices fit their circumstances, a particular risk where LLM recommendations may not account for local labor markets or infrastructure realities.

These concerns are evident in African-specific research as well. Despite the potential of generative AI in supporting career guidance, it can reproduce biases embedded within its training data, potentially misleading job seekers who accept AI advice without understanding its limitations. LLMs have also been connected to cultural bias, often creating responses that reflect Western norms while underrepresenting locally relevant perspectives (Ade-Ibijola & Okonkwo, 2023). In some instances, LLMs provide outputs that ignore the experiences of certain gender and minority groups, reinforcing the structural inequalities they are supposed to address (Ade-Ibijola & Okonkwo, 2023; Kotek et al., 2023). Consequently, the integration of LLMs in career guidance requires close oversight and sustained human involvement to ensure recommendations are accurate and reflective of local contexts.

## 2.2 AI Adoption and Computing Education in African Contexts

The patterns described above take on particular meaning when examined through an African lens. While students worldwide share similar concerns about AI in education (Ravšelj et al., 2025), the infrastructure, policy, and cultural contexts in African nations create distinct conditions that determine not just whether LLM-based career guidance is technically viable, but whether it can serve students well.

AI adoption across the continent has been met with varying degrees of readiness, with many African institutions held back by infrastructure and policy barriers that slow effective implementation (Eke et al., 2023). A growing "AI divide" has emerged, where wealthier economies capture most of the benefits from AI progress while emerging regions, particularly in Africa, struggle to keep pace (United Nations, 2024). This gap is evident in the 2024 Oxford Insights Government AI Readiness Index, which ranks Sub-Saharan Africa as the lowest-scoring region with an average readiness score of 32.70, though Data and Infrastructure was identified as its strongest area with a score of 42.06 (Oxford Insights, 2024). Egypt, Mauritius, South Africa, and Rwanda led Africa with scores slightly above 50, while countries like Senegal, Nigeria, Kenya, and Ghana also showed evidence of steady growth in AI readiness and capacity. Despite these relatively low rankings, AI adoption is steadily expanding across the continent, with visible applications in financial technology (fintech) and healthcare (Eke et al., 2023).

One dimension of the AI divide that existing career guidance research has overlooked is linguistic diversity. In many African contexts, computing course materials fail to reflect students' cultural and linguistic diversity, with challenges such as diverse dialects and direct translation difficulties requiring bilingual approaches (Dodoo et al., 2025). AfroBench, a benchmark for evaluating LLM performance on African languages, covers 64 languages across 15 tasks including natural language understanding, text generation, and question answering (Ojo et al., 2025). This evaluation shows large performance gaps between English and African languages, with performance varying based on available training data. The implications for career guidance are direct: if LLMs struggle with basic language tasks in African languages, their ability to provide nuanced, culturally appropriate career advice to students who think and communicate primarily in these languages is limited. Even when students interact with LLMs in English, they may be seeking guidance for careers in contexts where local languages dominate professional communication, a mismatch that current LLMs are not equipped to handle.

Beyond language, the broader picture of AI adoption in African higher education shows both opportunities and unresolved challenges. A review of AI adoption across five African universities, combining surveys of 230 faculty members with 24 interviews, identified benefits including personalized learning, administrative efficiency, and research enhancement (Maluleke, 2025). However, the review also illustrated challenges: digital divide concerns, data privacy issues, infrastructure gaps, and policy limitations. The study argued that context-specific governance frameworks and equitable infrastructure investment are needed before AI can be successfully integrated. What remains largely unexplored is how students engage with and experience AI tools for high-stakes decisions like career planning, or how guidance quality compares across different national contexts.

**Table 1: AI Readiness Performance Across Selected African Nations Source:** Oxford Insights (2024). Government AI Readiness Index 2024. *This table positions our ten selected countries (highlighted in bold) within the broader African AI readiness landscape*

| Rank (Africa) | Country | AI Readiness Score | Global Rank | Government | Technology Sector | Data & Infrastructure | Strongest Pillar |
|---|---|---|---|---|---|---|---|
| 1 | **Egypt** | 55.63 | 75 | 68.98 | 42.13 | 55.77 | Government (68.98) |
| 2 | Mauritius | 53.94 | 82 | 65.31 | 32.71 | 63.81 | Government (65.31) |
| 3 | **South Africa** | 52.91 | 85 | 54.30 | 39.15 | 65.28 | Data & Infrastructure (65.28) |
| 4 | Rwanda | 51.25 | 93 | 71.44 | 30.30 | 52.02 | Government (71.44) |
| 5 | **Senegal** | 46.11 | 108 | 62.37 | 28.77 | 47.18 | Government (62.37) |
| 6 | **Tunisia** | 43.68 | 114 | 28.62 | 41.07 | 61.35 | Data & Infrastructure (61.35) |
| 7 | **Kenya** | 43.56 | 115 | 56.20 | 30.98 | 43.49 | Government (56.20) |
| 8 | **Nigeria** | 43.33 | 116 | 59.88 | 27.11 | 42.99 | Government (59.88) |
| 9 | **Ghana** | 43.30 | 117 | 59.53 | 25.35 | 45.03 | Government (59.53) |
| 10 | **Benin** | 42.97 | 119 | 59.92 | 24.30 | 44.68 | Government (59.92) |
| 11 | **Zambia** | 41.87 | 123 | 60.78 | 23.22 | 41.63 | Government (60.78) |
| 12 | **Morocco** | 41.78 | 124 | 34.82 | 36.70 | 53.82 | Data & Infrastructure (53.82) |

As seen in Table 1, Egypt leads at 75th globally while Morocco ranks 124th, placing both nations in the global middle tier, well behind leaders like the US (1st) and Singapore (2nd), but representing Africa's most AI-ready economies. Notably, nine of the twelve countries achieved their highest scores in the Government pillar, suggesting that African nations are developing AI strategies and governance frameworks even as technological infrastructure lags. With Sub-Saharan Africa averaging just 32.70 compared to 70+ for high-readiness nations, this gap highlights a critical challenge: AI tools developed in technologically advanced contexts may embed assumptions about infrastructure and career pathways that do not reflect the realities of many African students, a key consideration for examining LLM-generated career guidance in these settings.

Underlying this AI readiness gap are fundamental educational challenges. While curricula in countries like Nigeria cover key technical areas like AI, cloud computing, cybersecurity, and data analytics, these subjects are frequently taught without practical exposure, limiting students' ability to translate theory into practice (NITDA, 2022). Evidence from the World Bank's Skills Toward Employment and Productivity (STEP) Employer Survey in Kenya found that firms face difficulties in recruiting because applicants often lack essential skills and experience (Sánchez Puerta et al., 2018). Employers consistently rank transversal skills like communication and problem-solving as critical yet underdeveloped among graduates. These gaps are worsened by limited access to computing tools, unreliable internet, outdated curricula, and weak academic-industry partnerships (Isuku, 2018).

The emergence of LLMs offers a promising alternative to close these gaps, given the ability of these systems to explain technical requirements on demand, help with problem-solving, and provide career advice even over basic internet connections (Hamouda et al., 2025; Xu et al., 2024). Research on the state of AI-driven career guidance in West Africa documents innovative infrastructural developments like solar-powered AI centers and SMS-based systems to help navigate infrastructural limitations to AI adoption, presenting opportunities for increased adoption of AI-driven career guidance tools (Eze & Alabi, 2025).

## 2.3 Computing Careers, and the Evolving Skills Landscape

Careers in computing stand out from many other professions because of the way entry into the field is evaluated. Employers usually expect applicants to go beyond listing academic qualifications by demonstrating competence in practice. This includes solving problems through code, reasoning about algorithms and data structures, and performing technical assessments that often form part of interviews (Aasheim et al., 2009; Lunn et al., 2022). At the same time, hiring decisions are rarely made on technical ability alone. Increasingly, employers also assess professional skills such as teamwork, communication, adaptability, and ethical awareness (NACE, 2024; Africa Career Networks, 2023). These expectations reflect how computing roles now extend across multiple sectors, including healthcare, finance, education, government, and the creative industries (Wong, 2024).

The challenge for educators and administrators in Africa is figuring out how to prepare graduates for both sets of demands while dealing with limited resources and structural barriers (Eze et al., 2025). Some recent work offers potential solutions. AI-driven platforms can give students better access to career information and guidance, helping them understand what different sectors require (Asmamaw & Nsabiyumva, 2025). Similarly, AI tools like mobile coding tutors and bilingual teaching assistants can help students build their technical skills while filling gaps in access to quality instruction (Boateng, 2024). Both findings suggest that AI could play an important role in preparing students for emerging computing jobs in areas like data science, cybersecurity, and AI development.

Global trends add another layer to this picture. Sustained growth in AI-related computing jobs is projected worldwide, but persistent digital skills gaps could limit participation (World Economic Forum, 2023). Generative AI is also beginning to reshape "cognitive work," replacing some routine tasks while placing more value on reasoning, collaboration, and continuous learning (Brookings Institution, 2024). For computing graduates in Africa, this makes adaptability and breadth of skills just as important as technical competence. Building AI-related content into curricula and aligning academic training with industry expectations can help ensure that students graduate with the tools needed to succeed in a changing employment landscape.

## 2.4 Research Gaps and Study Rationale

Despite growing interest in AI applications for education, critical research gaps persist. Most studies on AI adoption in career guidance focus exclusively on Western contexts, creating knowledge gaps about how these tools are used within African educational contexts where unique cultural epistemologies, multilingual requirements, and institutional frameworks shape computing career pathways (Brandenburger et al 2023; Leung 2022). This geographic bias perpetuates a form of technological colonialism in educational AI, where Western-developed solutions are universally applied without consideration of local educational values (Santos, 2007). Two recent studies have started to address this gap, but leave important questions open. Research on AI challenges in African education and highlighted cultural bias and the need for African-centered approaches, though without empirically comparing how different LLMs perform when generating career guidance or examine variation across countries with different educational systems and labor markets (Ade-Ibijola & Okonkwo, 2023). AI applications in educational contexts have been examined with attention to implementation challenges, though without a standardized framework for evaluating career guidance quality or comparing outputs across models (Hamouda et al., 2025). Neither study compared proprietary versus open-source models, a distinction that matters for resource-constrained institutions.

This study addresses these gaps by providing the first multi-country comparison of LLM-generated career guidance, using a framework that blends CC2020 competencies with cultural responsiveness measures to assess how different LLM architectures handle technical, professional, and contextual dimensions across ten African nations

## 3. Theoretical and Analytical Frameworks

### 3.1 Theoretical Framework: Digital Colonialism and Ubuntu Philosophy

This study draws on two complementary theoretical perspectives to interpret the role of LLMs in African education: Digital Colonialism Theory and Ubuntu Philosophy. These frameworks guided the analysis of our findings by exposing the underlying power asymmetries in global technology development and deployment, while offering a human-centered, culturally grounded alternative for interpreting AI integration in African contexts. We elaborate on each separately before describing how they may fit together as a conceptual model to shape this inquiry.

**Digital Colonialism Theory:** Digital Colonialism theory explains how powerful nations and corporations from the Global North exert control and extract value from the Global South through digital technologies and infrastructures in ways that resemble historical forms of colonialism. Specifically, it involves control over digital spaces, data, and other technological platforms to both overtly and covertly exert political, economic, and social control over less powerful nations or groups (Kwet, 2019; Singler & Babalola, 2024). Previously, Kwet (2019) argued that the dominance of Western-based tech companies in Africa's digital infrastructure and content represents a modern form of colonialism, creating dependence on foreign-owned platforms, software, and services. Coleman (2019) described this as the new scramble for Africa, where technologically advanced countries shape the trajectories of African nations to suit their own needs through data and information technologies. In the context of African education, this has far-reaching implications, with this form of technological control contributing to deepening global inequalities, allowing external actors to influence how knowledge is produced, accessed, and disseminated across African institutions (Kwet, 2019).

This theoretical underpinning offers a useful lens for examining the non-local perspectives often present in AI tools such as LLMs, primarily trained on Western data sources. As a result, these systems frequently reflect foreign thought patterns and built-in assumptions that may not fit local realities in Africa. Digital Colonialism Theory helps to explain how such biases, infrastructural presumptions, and standardized recommendations can reflect the continuation of colonial patterns of dominance in the contemporary digital space (Birhane, 2020; Mohamed et al., 2020).

**Ubuntu Philosophy:** In contrast, the African philosophy of Ubuntu presents a decolonial, human-centered approach grounded in values of collectivism, mutual care, and context-based understanding. Captured in the Nguni Bantu phrase "*umuntu ngumuntu ngabantu*" (a person is a person through other persons), Ubuntu holds that individual well-being and success are deeply connected to the well-being of the community. It also views knowledge as something that is co-created through shared experience, relationships, and collective values (Letseka, 2013). In some African educational settings, Ubuntu has been embraced as a guiding philosophy that values indigenous ways of learning, communal knowledge-sharing, and the integration of ethical and spiritual dimensions into education (Omodan & Diko, 2021; Waghid, 2013)

While Ubuntu is one tradition among many across the continent, we draw on it here because it offers specific evaluative criteria, such as collective well-being, community involvement, and relational success, that contrast with the individualistic assumptions often embedded in conventional AI systems. When applied to career guidance, an Ubuntu-based approach would emphasize community involvement over a one-size-fits-all approach, considering a learner's social background, local mentorship networks, and alignment with collective goals, a sharp contrast to the individualistic and market-driven ethos that often guides Western technological approaches. Hence, by grounding our analysis in Ubuntu, this study advances the need for technologies created with and for African communities, while also examining the current limitations of LLMs in providing career guidance that is contextually relevant for African learners.

Digital Colonialism Theory helps explain why LLMs trained on Western data might assume particular infrastructures, career pathways, and professional norms that do not fit African contexts, but it stops short of offering alternative standards for what locally relevant guidance should include. Ubuntu Philosophy fills this gap by foregrounding collective well-being, community involvement, and relational success as criteria against which AI recommendations can be evaluated. Their combination allows us to both identify where current LLM outputs fall short and assess how closely they align with values that matter in African educational settings.

### 3.2 Analytical Framework: Computing Curricula 2020

This study adopts the CC2020 framework as its primary analytical lens. Unlike alternatives such as the Skills Framework for the Information Age (2018) or the European e-Competence Framework (Comité Européen de Normalisation, 2018), CC2020 provides the most comprehensive and widely recognized guide for computing education globally, encompassing both the technical and workplace skills needed in today's computing jobs (Clear et al., 2020).

Constructivist learning theory (Piaget, 1977) views knowledge as something people construct through active interaction with their environment and forms the epistemological basis of this study. This viewpoint helps us understand how LLMs provide career advice and how different contexts shape their responses. We use CC2020 not as a theoretical foundation but as an analytical tool for organizing and interpreting data, while still allowing space for new patterns and discoveries to emerge (Ravitch & Riggan, 2012). This approach gives us both theoretical depth and methodological clarity, helping to address common concerns about a lack of structure in newer studies on AI and education.

### 3.2.1 Operationalization of CC2020 Competency Dimensions

The CC2020 framework divides computing competencies into three areas: knowledge, skills, and dispositions. We used these categories to examine the LLM responses, enabling the outputs to be matched with the relevant parts of the framework.

**Knowledge:** This covers subject-based content, including programming, algorithms, software engineering, systems architecture, artificial intelligence, cybersecurity, and data science. We broke these down into specific areas in some instances to better assess how completely the models covered each topic.

**Skills:** This component focuses on how the models demonstrate practical abilities, including problem-solving, computational thinking, design thinking, project planning, effective communication in technical settings, and collaboration. These skills represent the link between what people know and how they apply it in the workplace.

**Professional Dispositions:** This encompasses attitudinal and behavioral competencies, including ethical awareness in the context of technology use, commitment to lifelong learning, cultural competence for global technology work, a collaborative mindset, social responsibility in computing applications, and professional accountability. This part covers the human aspects of computing careers that often get overlooked due to emphasis on technical skills or competency.

Section 4 details how these CC2020 dimensions worked alongside our theoretical framework during coding.

### 3.3 Framework Integration and Conceptual Model

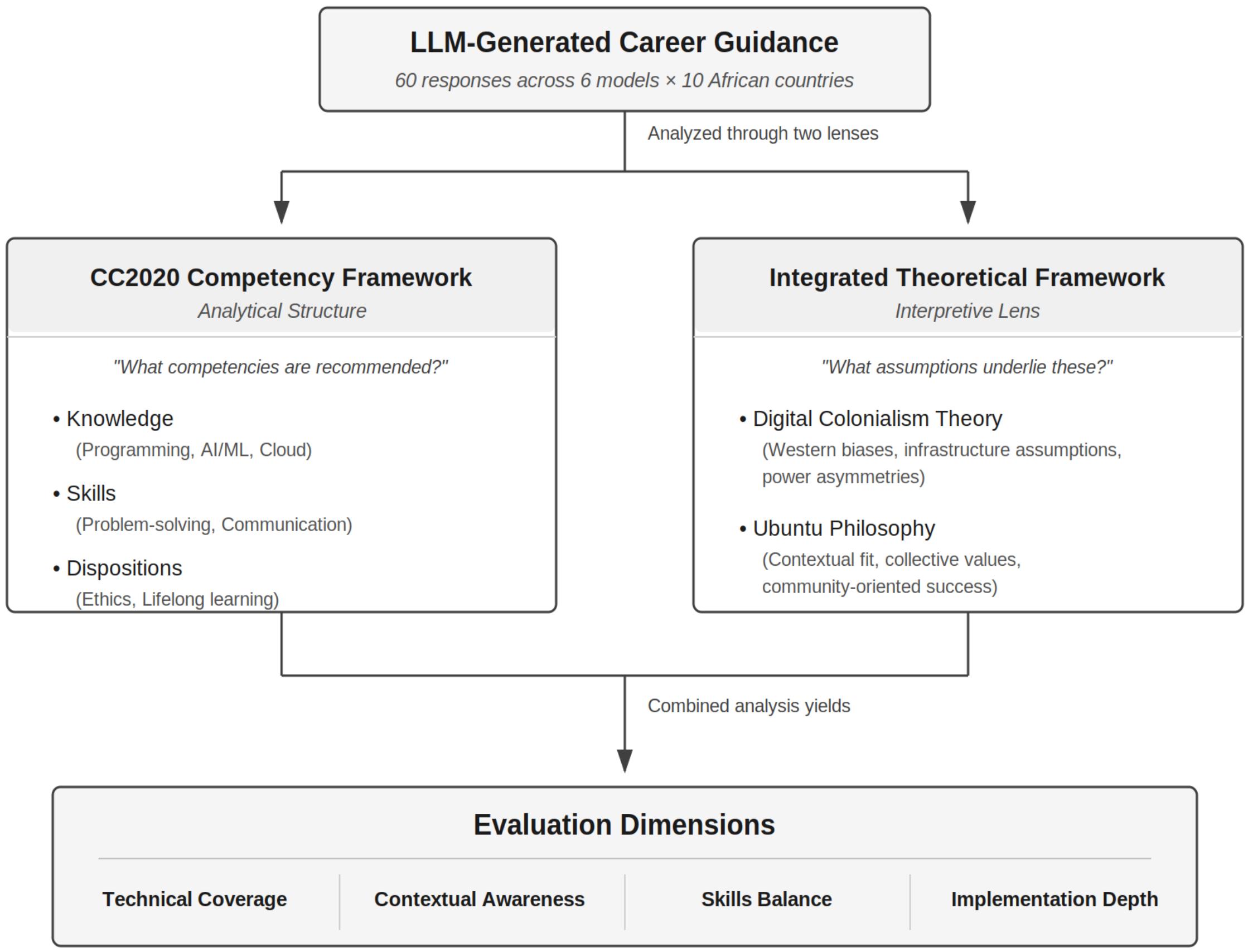

**Figure 1.** Theoretical Framework for Interpreting AI-Generated Career Guidance in African Contexts

Having established our theoretical foundation and analytical structure, we now explain how these three components work together in practice. The CC2020 framework provides the structure for categorizing computing competencies, while our theoretical perspective determines what we look for within these categories and how we interpret findings. This means using CC2020 to code LLM responses while applying our theoretical framework to question the assumptions behind these recommendations. For example, when an LLM suggests specific technical competencies, CC2020 helps us categorize them properly, but Digital Colonialism Theory makes us ask: whose infrastructure is assumed? When professional competencies appear, Ubuntu Philosophy guides us to examine whether they reflect individual achievement or collective success models.

This combined approach directly addresses our RQs. For RQ1 and RQ2 about similarities and differences across countries, CC2020 provides the consistent categorization needed for comparison. Our theoretical perspective then, is applied to explain why certain competencies might appear across contexts, whether this reflects genuine cross-national relevance or the reproduction of Western templates that Digital Colonialism Theory would predict. For RQ3, which compares LLM performance, the integration of all three frameworks shapes our evaluation approach: examining technical and professional competency coverage through the CC2020 lens, assessing contextual awareness through the perspectives of Ubuntu philosophy and Digital Colonialism Theory, and evaluating the overall depth and relevance of recommendations for African computing graduates. Through this integration, we move from theory to structured evaluation, positioning our analysis to uncover both current limitations and future possibilities for AI-driven career guidance in African contexts.

## 4. Methodology

### 4.1 Research Design

Content analysis was selected as the methodology given its focus on examining textual data systematically to allow for both quantitative assessment and qualitative interpretation of LLM outputs (Hsieh & Shannon, 2005; Krippendorff, 2018). Following Mayring's (2014) methodology, we implemented a directed content analysis approach, where the CC2020 framework provided initial coding categories. This approach enabled structured examination of textual data across multiple variables (different LLMs and country contexts), providing a replicable methodology for evaluating AI-generated content while allowing for emergent themes. The directed approach guides consistent coding while maintaining sensitivity to patterns not anticipated by the initial framework. Figure 2 provides an overview of the six-stage methodology, which is detailed in the following subsections.

To ensure that each output was generated based on the same criteria, a single standardized prompt was employed for all model-country queries. Each LLM was given the exact same query about entry-level computing career advice in a specified country, facilitating direct comparison across outputs. However, this controlled prompt strategy limited the opportunity for models to demonstrate adaptive dialogue or clarifying questions, a trade-off we acknowledge in our limitations. As an exploratory descriptive study, this approach is appropriate for answering our research questions, which seek to characterize LLM outputs and identify patterns across contexts rather than test causal relationships between models.

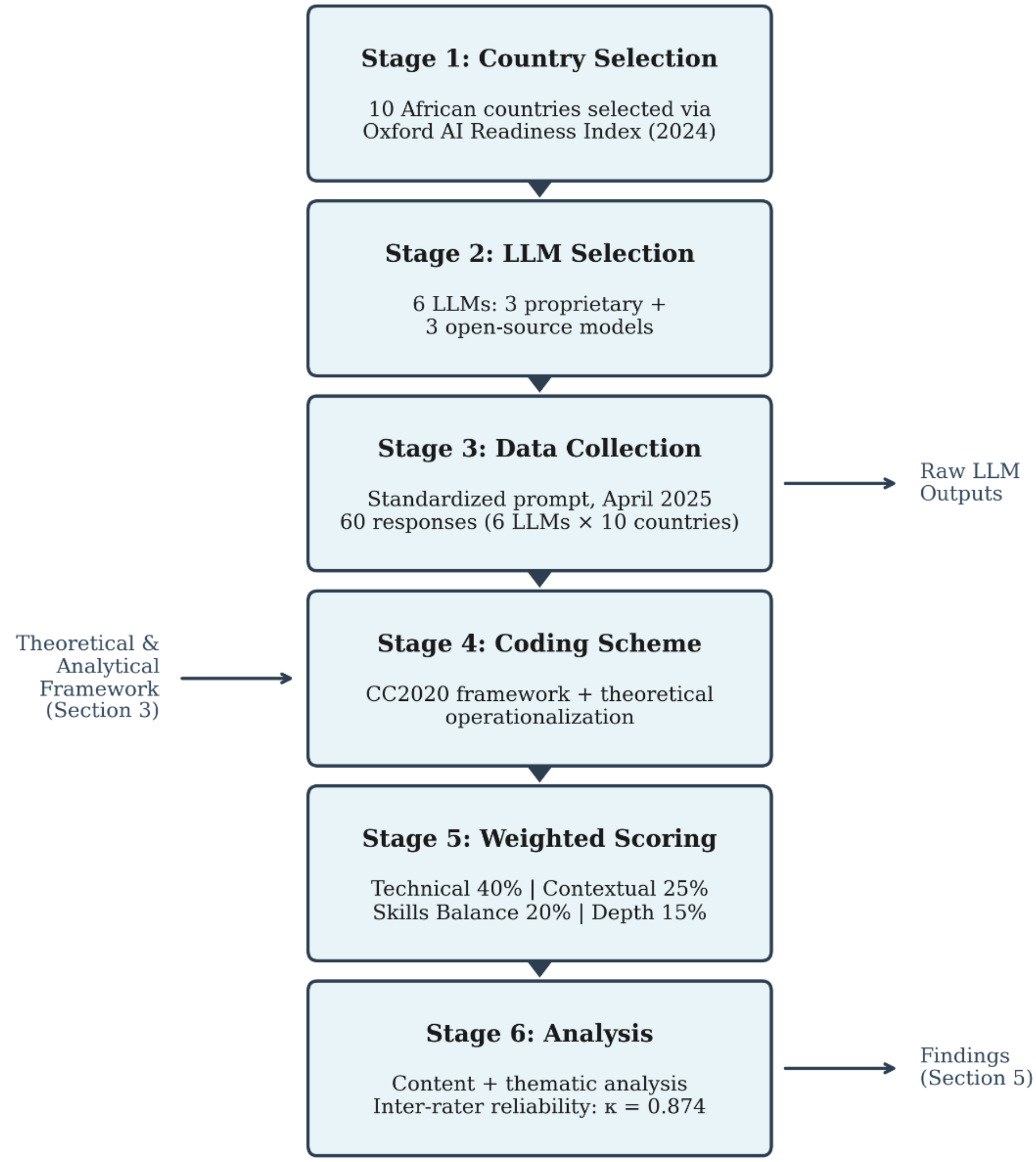

**Figure 2.** Overview of the six-stage research methodology for evaluating LLM-generated career guidance across ten African countries.

## 4.2 LLM Selection

Six LLMs were selected to represent the current landscape of accessible AI tools, balancing proprietary and open-source options. This distinction proves relevant for resource-constrained contexts. Proprietary models typically require paid subscriptions ($20-30/month for individual access), while open-source alternatives can be accessed freely through web interfaces or deployed locally without licensing fees, offering greater accessibility for institutions with limited budgets. **Proprietary models** are software products owned and controlled by specific companies, where the source code remains closed and access is typically provided through paid licenses or subscriptions with restricted modification rights (Raymond, 2001; Stallman & Gay, 2002). The proprietary models selected included ChatGPT-4 (OpenAI), widely adopted in educational

contexts with strong natural language generation capabilities; Claude 3.5 "Sonnet" (Anthropic) known for nuanced responses and a design focus on ethical considerations; and Gemini (Google) integrated with Google's ecosystem and optimized for factual accuracy.

**Open-source models**, in contrast, are distributed with publicly accessible source code under licenses that permit users to view, modify, and redistribute the software, often developed through collaborative community-driven processes (Raymond, 2001). The open-source models examined included Llama 3 (Meta), representative of high-quality open-source alternatives; DeepSeek-V2 (DeepSeek AI), an emerging model with strong technical performance and specialized training focus; and Mistral 7B (Mistral AI), an efficient, smaller-scale model.

All models were accessed through their respective web interfaces using default settings. The versions available at data collection were: ChatGPT (GPT-4 via chatgpt.com), Claude 3.5 Sonnet (via claude.ai), Gemini 1.5 Pro (via gemini.google.com), Llama 3 70B (via meta.ai), DeepSeek-V2 (via chat.deepseek.com), and Mistral (via chat.mistral.ai). All LLM queries were conducted in English to control for language variation and because English serves as a common language of instruction or professional communication in the selected countries. We acknowledge that this English-only approach represents a limitation given Africa's multilingual reality, but it enabled focus on content differences beyond basic translation considerations. For each country-LLM combination (6 models × 10 countries = 60 outputs), we employed the standardized prompt: *"What LLM skills or competencies do students applying for a computing job in [Country] need to know and how should they prepare for an entry-level position?"* While this phrasing includes the term 'LLMs,' our focus was not on LLM-specific skills, but on the broader career guidance generated by the models. We therefore interpret the responses as reflecting general career competencies, consistent with our study's aims.

Data collection occurred within a 48-hour period in April 2025 to minimize potential effects of model updates. A new account was created for each model, and each country query was the first prompt submitted, ensuring no prior interaction history could influence the outputs. Single queries were used for each country-model combination, as pilot testing showed consistent responses to identical prompts across multiple runs.

We used single-run prompts rather than multiple trials for two reasons. First, preliminary testing showed that responses to identical prompts were consistent when submitted multiple times within short intervals, suggesting response variability was minimal under controlled conditions. Second, the exploratory nature of this study prioritized breadth of coverage (six models across ten countries) over depth of repeated sampling, a trade-off we acknowledge in our limitations. The resulting text responses were organized by model and country for subsequent analysis. No human interaction beyond the initial prompt was permitted; we refrained from follow-up questions or response clarification to evaluate each model's immediate guidance capabilities. All LLM responses, coding sheet, and scoring tables are available as supplementary materials to provide transparency and enable replication.

### 4.3 Coding Scheme Development and Scoring Methodology

We employed a structured coding procedure to analyze the LLM responses using the CC2020 framework (as detailed in section 3.2), which includes major technical domains (e.g., programming, data science, networking) and professional skill areas (Clear et al., 2020). This provided an initial deductive codebook covering expected computing competencies for entry-level positions.

Alongside CC2020, our theoretical framework shaped specific coding decisions. Digital Colonialism Theory directed attention to instances where LLM responses assumed Western infrastructure (e.g., specific cloud platforms, software ecosystems), recommended career pathways typical of Silicon Valley contexts, or provided standardized advice without acknowledging local conditions. Ubuntu Philosophy led us to examine whether responses referenced community networks, collective success, local mentorship, or alignment with communal values (e.g., suggesting graduates contribute to local tech communities or mentor peers) rather than purely individualistic career advancement. These theoretical indicators complemented the CC2020 categories, allowing us to assess not only what competencies were mentioned but also whether the underlying assumptions fit African contexts.

We then conducted a pilot coding on a subset of responses (one country across all models) to refine the scheme and ensure comprehensive coverage. During this process, we remained attentive to any new themes or competencies not encompassed by CC2020. Where inductive themes emerged (for example, references to specific local technologies, cultural practices, or regulatory frameworks), we adjusted the codes accordingly, ensuring the coding scheme captured both predetermined competency areas and emergent content specific to African contexts. The pilot coding revealed several areas requiring refinement. Initial categories for technical competencies were too broad, leading us to distinguish between general programming skills and context-specific technical requirements (e.g., mobile development for markets with high smartphone penetration). We also added codes for infrastructural assumptions after noticing that some models recommended tools requiring stable internet connectivity without acknowledging potential access limitations. These refinements illustrate how the coding scheme evolved through engagement with the data rather than being rigidly imposed from the outset. Throughout this process, coding focused on conceptual meaning rather than strict keyword matching; responses were coded based on the presence of an idea even if exact terminology varied (e.g., 'privacy' or 'compliance with regulations' was coded under ethics even if the word 'ethics' itself was not used).

**Contextual Awareness Scoring (0-4 Scale):** We measured contextual awareness using an additive scoring system across four dimensions, with each dimension scored as present (1) or absent (0), based on the "matrix of evidence for validity argumentation protocols" for judging cultural responsiveness in educational technology research (Solano-Flores, 2019). We gave one point for each of these features that appeared in a response: (1) reference to the local tech industry,

(2) attention to language or cultural factors, (3) mention of national policies or regulations, and (4) reference to local institutions or education programs. This additive method follows standard practice in educational assessment, where overall scores show the accumulation of demonstrated competencies (Williamson et al., 2012).

It is important to distinguish between two measures of contextual awareness used in this study. The first is an additive indicator score (0-4), which counts the presence or absence of four contextual elements: local tech industry reference, language/cultural considerations, national policy mention, and local institution reference. This measure, reported in Section 5.2, captures whether specific contextual elements were present. The second is a rubric-based contextual awareness score (0-5), which evaluates the quality and depth of contextualization as part of the composite performance framework. This measure, weighted at 25% in the composite score, reflects holistic judgment of how well a response adapted to local context. The complete scoring rubric is provided in the codebook (Appendix B).

When responses contained ambiguous or borderline content, we applied a conservative coding approach. For contextual awareness, a reference was coded as present only if it named a specific local entity, policy, or practice rather than making generic statements about "local conditions." For example, mentioning "Nigeria's NITDA guidelines" counted as a policy reference, while stating that graduates should "be aware of local regulations" did not. Similarly, technical competencies were coded based on explicit mention rather than implied coverage. Where coders disagreed on borderline cases, they discussed the specific passage and reached consensus based on these operational definitions.

**Weighted Evaluation Criteria Justification:** The scoring weights (Technical Coverage 40%, Contextual Awareness 25%, Skills Balance 20%, and Implementation Depth 15%) are grounded in the CC2020 competency framework and research on computing graduate employability. CC2020 frames computing competency as the integration of knowledge, skills, and dispositions enacted in practice, where disciplinary knowledge ("knowing what") provides a necessary foundation complemented by skilled application ("knowing how") and professional orientation ("knowing why") (Clear et al., 2020). Employer research consistently illustrated that while interpersonal and professional attributes are highly valued, technical proficiency functions as a baseline requirement for entry-level consideration; employers expect graduates to arrive with core technical competencies already in place (Aasheim et al., 2009; Harvey, 2005), and structured technical preparation influences graduates' ability to secure computing positions and achieve favorable early labor market outcomes (Mason et al., 2009). Contextual awareness receives substantial weight (25%) because CC2020 emphasizes that global competency frameworks must be interpreted in relation to institutional, national, and cultural conditions rather than applied uniformly (Section 7.1.7, p. 81), and research on culturally responsive assessment demonstrates that local relevance is central to the validity of educational tools across diverse settings (Squires & Preece, 1999; Solano-Flores, 2019). Skills Balance (20%) reflects CC2020's position that technical knowledge alone is insufficient for professional preparation and aligns with career-readiness frameworks defining employability through the integration of technical and transferable competencies (Clear et al., 2020; NACE, 2024). Implementation Depth (15%) is justified by CC2020's characterization of skill development as requiring deliberate practice and engagement with progressively higher-order cognitive processes, making specificity and actionability meaningful indicators of applied competence. The proportional allocation follows established practices in weighted rubric design, where criteria are assigned differential weights according to their relative importance to the construct being evaluated (Allen & Tanner, 2006; Williamson et al., 2012). Initial weights derived from this literature were refined through pilot coding to ensure the scheme appropriately differentiated model performance. We acknowledge that alternative weightings could yield different results. The composite scores are presented as heuristic indicators of relative performance rather than precise measurements, and readers should interpret rank orderings with appropriate caution, given the exploratory nature of this study.

**Implementation Depth Operationalization:** Implementation Depth was measured using a 0–5 scale to evaluate how specific and actionable each LLM recommendation was, following guidelines used in the evaluation of automated scoring systems (Williamson et al., 2012). A score of 5 was given to responses that mentioned some very relevant and specific tools, clear learning pathways, and concrete, actionable steps. Scores of 3–4 reflected general advice with some specific or partially actionable content. Responses that were vague, overly broad, or lacked clear direction received scores of 1–2. A score of 0 was assigned when no meaningful implementation guidance was present.

The thematic analysis followed Fereday and Muir-Cochrane's (2006) hybrid approach, moving from coding, theme development, and final refinement. This enabled combining deductive coding using the CC2020 framework with inductive identification of emergent themes. Each criterion was evaluated using a consistent 5-point scale, with clearly defined benchmarks to guide scoring.

Inter-rater reliability was established through independent blind double-coding of all 60 responses by the first two authors, meaning each coder worked without knowledge of the other's decisions until comparison. Across 1,200 binary coding decisions, pooled agreement between the two ratings was high ($\kappa = 0.874$, 93.8% observed agreement), indicating substantial to almost perfect consistency. Reliability was strongest for professional skills ($\kappa = 0.955$), followed by technical competencies ($\kappa = 0.774$) and contextual factors ($\kappa = 0.756$). Where disagreements arose, the first two authors discussed the specific passages and reached consensus based on the operational definitions established during pilot coding. This process resolved all discrepancies without requiring third-party adjudication. To further mitigate potential bias, the lead researcher engaged in peer debriefing sessions with a specialist in computing education who reviewed a sample of coded responses and provided feedback on interpretive consistency. These steps provide confidence in the validity and reliability of the subsequent analyses.

The complete codebook, including definitions for all codes, indicators, and sample coded excerpts, is available in the study's data repository (Eze & Lunn, 2026). Appendix A contains example prompts and representative LLM outputs to illustrate the data and coding process.

### 4.4 Researcher Positionality

The first author led the study design, analysis, interpretation, and writing of the manuscript. He identifies as a Nigerian researcher focused on AI-driven career guidance and contextually relevant educational technologies for African learners. He recognizes that his academic trajectory from Chemical Engineering in Nigeria and South Africa to AI research in the United States shapes how he approached this investigation. His work with the Nigerian Artificial Intelligence Research Scheme and the Savannah Digital Research Institute on AI-driven career guidance has illuminated both the promise and limitations of these technologies in African contexts. While he believes strongly in AI's potential to augment career guidance, he maintains that these tools should complement, not replace, human counselors, and must reflect Ubuntu principles of collective success. This dual perspective shaped his analysis throughout the study, aiming to highlight the diverse needs of African computing students while taking a critical stance on how mainstream AI tools serve these populations.

The second author contributed to the study design, analysis, and critical review of the manuscript. She identifies as an American assistant professor, whose research and teaching are centered on fostering technical and professional competencies in tertiary computing students. In particular, she considers this with regard to preparation for the workforce. Although she is not from the Global South, she has previously explored bias around LLMs and seeks to work towards more inclusive consideration of training data and the potential implications of these tools. She has also created and led a course on AI, ethics, and societal impact. Her experiences and background influenced her perspective and the emphasis on the potential benefits of LLMs for career guidance within more specific contexts, while also seeking to promote the value of responsible use of these technologies.
The third author offered a high-level review of the paper after the first and second authors completed the manuscript. He is an associate professor and identifies as a 2.5 generation Eritrean American. In addition to this study, he is currently exploring how AI tools can inform the pathways of prospective STEM transfer students who begin their careers in U.S. community colleges. Similar to the first author, he believes that this transfer tool and other academic/career guidance platforms should complement rather than replace traditional advising. This perspective is largely informed by his prior career in an engineering student affairs role, in which he worked closely with academic advisors across eight different departments.

## 5. Results

The following sections present the findings from our content analysis. Section 5.1 examines how the models covered technical and professional competencies, finding both areas of convergence and divergence. Section 5.2 examines the differences in contextual awareness and local adaptation. Section 5.3 evaluates overall model performance using our weighted scoring framework. Together, these analyses address how different LLMs conceptualize computing competencies for African contexts and their relative effectiveness as career guidance tools.

### 5.1 Technical and Professional Competency Coverage

Analysis of the 60 LLM responses demonstrated differences in how models balanced technical and professional competencies for entry-level computing jobs across the ten countries. As summarized in Table 2, technical competencies dominated overall, with programming, Python, and AI/Machine Learning (ML)/Natural Language Processing (NLP) consistently highlighted. Professional competencies such as adaptability, teamwork, and communication also appeared in every model output, though emphasis varied. In contrast, specialized areas like cybersecurity, data analytics, and project management appeared much less frequently, suggesting gaps in model-generated guidance for certain career pathways.

**Table 2.** Frequency of Technical and Professional Competencies Mentioned Across Six LLMs (N=60 Responses)

| Code/Theme | ChatGPT-4 | Claude | Gemini | Mistral | Llama | DeepSeek | Total | Coverage |
|---|---|---|---|---|---|---|---|---|
| **TECHNICAL COMPETENCIES** | | | | | | | | |
| ***Core Programming*** | | | | | | | | |
| Programming | 41 | 19 | 28 | 22 | 41 | 30 | 181 | 100% |
| Python | 11 | 16 | 10 | 11 | 10 | 20 | 78 | 100% |
| Algorithms & Data Structures | 5 | 1 | 28 | 11 | 130 | 29 | 204 | 100% |
| ***Applied Domains*** | | | | | | | | |
| AI / ML / NLP | 156 | 37 | 164 | 176 | 90 | 271 | 894 | 100% |
| Cloud Computing | 69 | 20 | 0 | 77 | 50 | 70 | 286 | 83.3% |
| Databases & SQL | 15 | 2 | 12 | 0 | 50 | 3 | 82 | 83.3% |
| Software Engineering & Version Control | 12 | 9 | 31 | 0 | 60 | 49 | 161 | 83.3% |
| ***Specialized Areas*** | | | | | | | | |
| Cybersecurity | 10 | 0 | 0 | 0 | 10 | 0 | 20 | 33.3% |
| Data Analytics | 17 | 0 | 9 | 0 | 0 | 0 | 26 | 33.3% |
| Java | 6 | 0 | 7 | 0 | 10 | 9 | 32 | 66.7% |
| **PROFESSIONAL COMPETENCIES** | | | | | | | | |
| ***Core Employability Skills*** | | | | | | | | |
| Adaptability & Lifelong Learning | 28 | 3 | 28 | 11 | 20 | 22 | 112 | 100% |
| Teamwork / Collaboration | 9 | 4 | 22 | 11 | 35 | 37 | 118 | 100% |
| Communication | 8 | 1 | 21 | 11 | 12 | 10 | 63 | 100% |
| Problem-solving | 7 | 0 | 25 | 11 | 20 | 32 | 95 | 83.3% |
| ***Ethics & Management*** | | | | | | | | |

| Ethics & Responsible AI use | 38 | 7 | 9 | 0 | 0 | 42 | 96 | 66.7% |
|---|---|---|---|---|---|---|---|---|
| Project Management | 2 | 0 | 0 | 0 | 0 | 9 | 11 | 33.3% |

*Note: Total column represents cumulative frequency across all models. Coverage indicates the proportion of models (out of 6) that mentioned each competency at least once. Frequency counts reflect total mentions across all 60 responses.*

Several patterns emerge from the data. Core programming competencies (programming, Python, algorithms, and data structures) and applied domains (AI/ML/NLP, cloud computing, databases, software engineering) showed strong coverage, with most appearing in 83% to 100% of model outputs. The same held for core employability skills like adaptability, teamwork, communication, and problem-solving, which were broadly consistent across models. By contrast, specialized technical areas showed much lower and more uneven coverage. Cybersecurity appeared in only 33% of models, data analytics in 33%, and Java in 67%. Similarly, ethics and responsible AI use appeared in 67% of models but with notable variation; DeepSeek and ChatGPT-4 accounted for the majority of mentions, while Llama and Mistral included none. Project management was the least covered competency overall, appearing in only 33% of models. These competency patterns held across all ten countries, with models recommending similar technical and professional skills regardless of national context.

### 5.2 Contextual Awareness Analysis

Significant variation was observed in how the models tailored their responses to each country's context. As defined earlier in the methodology, we assessed contextual awareness across four dimensions and scored each response on a 0-4 scale. Table 3 summarizes the highest-scoring model for each country along with the contextual elements present. Figure 3 visualizes the complete indicator matrix across all 60 model–country responses, with detailed values provided in the supplementary material.

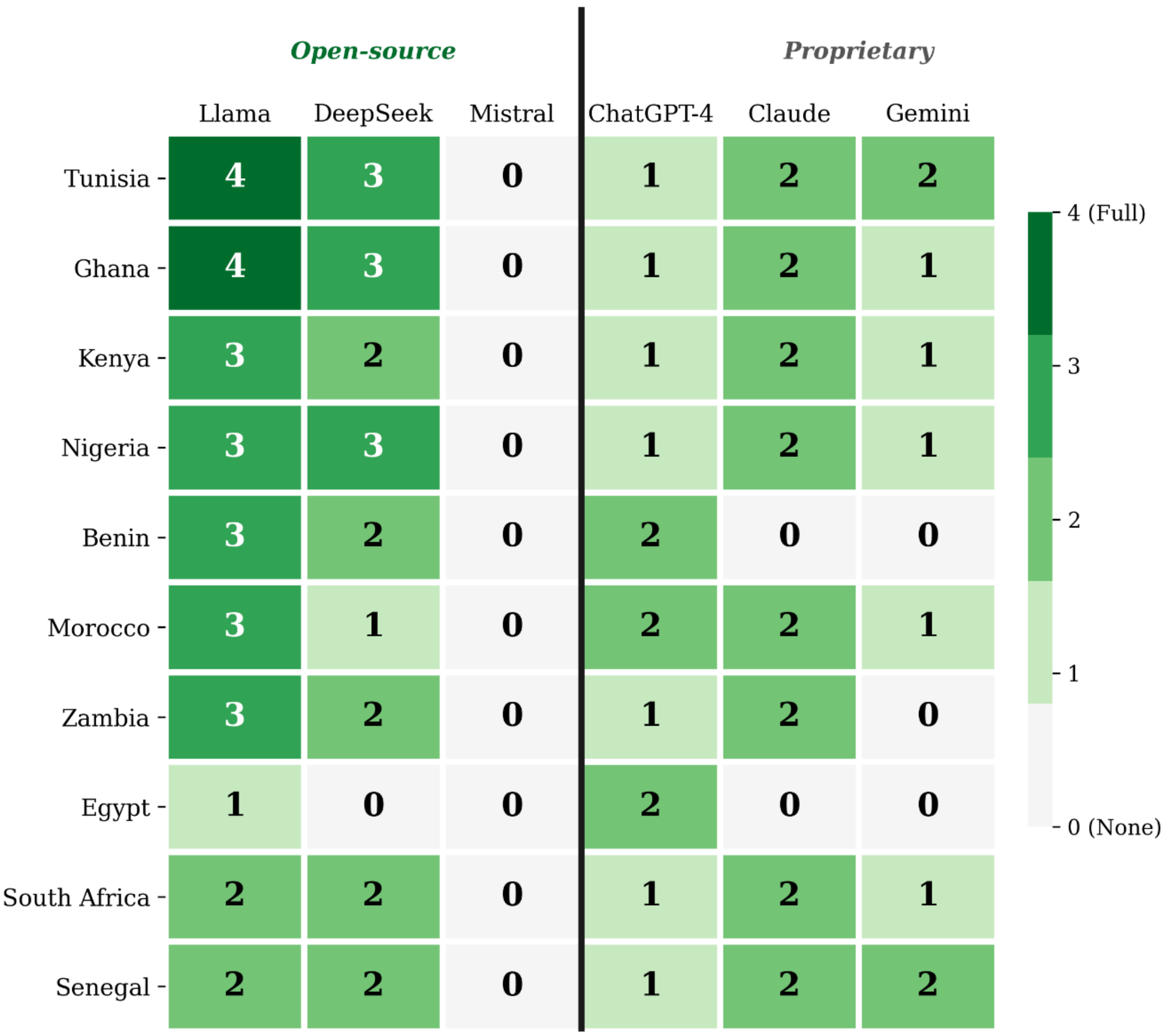

**Figure 3.** Contextual awareness scores by country and model. Scores represent the number of contextual elements present (0–4): local tech industry, language/culture, national policy, and local institution. Models are grouped by development type (open-source vs. proprietary).

**Table 3. LLM Contextual Awareness Performance by Country**

| Country | Contextual Awareness Score | Best Performing Model | Local Tech Industry | Language | National Policy | Local Institution | Key Contextual Elements Present |
|---|---|---|---|---|---|---|---|
| **Tunisia** | 4/4 | Llama | Yes | Yes | Yes | Yes | Tech hubs, Arabic/French proficiency, Digital Tunisia 2020, training programs |
| **Ghana** | 4/4 | Llama | Yes | Yes | Yes | Yes | Mobile money systems, tech ecosystem, government initiatives, education programs |
| **Kenya** | 3/4 | Llama* | Yes | Yes | No | Yes | M-Pesa, Silicon Savannah, fintech sector, local organizations |
| **Nigeria** | 3/4 | Llama* | Yes | Yes | No | Yes | Fintech industry, tech hubs, English proficiency, training institutions |
| **Benin** | 3/4 | Llama | Yes | Yes | Yes | No | Digital Benin strategy, French/English requirements, tech initiatives |
| **Morocco** | 3/4 | Llama | Yes | Yes | Yes | No | Digital Morocco 2025, Arabic/French context, innovation hubs |
| **Zambia** | 3/4 | ChatGPT-4* | Yes | Yes | No | Yes | Tech ecosystem development, English requirements, training programs |
| **Egypt** | 2/4 | ChatGPT-4 | Yes | Yes | No | No | AI/tech industry growth, Arabic/English bilingual requirements |
| **South Africa** | 2/4 | Claude* | Yes | Yes | No | No | Fintech sector, multilingual context, tech hub references |
| **Senegal** | 2/4 | Claude* | Yes | Yes | No | No | Dakar tech hubs, French/English proficiency requirements |

Scores represent the total number of contextual elements identified in each response, with a maximum of four. In some cases, multiple models achieved the same score, indicating tied performance and denoted with an asterisk (e.g., Nigeria: Llama and DeepSeek; South Africa: Claude and DeepSeek; Senegal: Claude, Gemini, Llama, and DeepSeek).

Of the 240 possible indicators across all 60 responses, only 85 were present, yielding an overall contextual awareness rate of 35.4%. Table 3a summarizes this measure by model. Llama achieved the highest contextual awareness at 67.5%, with at least one local reference in every response and perfect scores in Tunisia and Ghana, where its outputs mentioned national digital strategies like Digital Tunisia 2020, referenced local tech ecosystems, and acknowledged institutional training programs. DeepSeek followed at 50.0%, consistently attending to tech industry developments and language considerations, though it rarely mentioned national policies. Claude showed moderate performance at 40.0% but with notable gaps in countries like Egypt and Benin, while ChatGPT-4 reached 32.5% and Gemini 22.5%. Mistral, despite being open-source, offered no country-specific references in any of its 10 responses. Across all models, 15 of 60 responses contained no contextual indicators at all, with these gaps concentrated primarily among Mistral, Gemini, and Claude.

**Table 3a.** Contextual Awareness Summary by Model

| Model | Indicators Present (of 40) | Mean Score (0-4) | Awareness (%) |
|---|---|---|---|
| Llama | 27 | 2.70 | 67.5 |
| DeepSeek | 20 | 2.00 | 50.0 |
| Claude | 16 | 1.60 | 40.0 |
| ChatGPT-4 | 13 | 1.30 | 32.5 |
| Gemini | 9 | 0.90 | 22.5 |
| Mistral | 0 | 0.00 | 0.0 |
| **Overall** | **85** | **1.42** | **35.4** |

*Note: Each model evaluated across 10 countries x 4 indicators = 40 maximum. Awareness (%) = (indicators present / 40) x 100.*

### 5.3 Overall Performance Evaluation

Using the weighted evaluation framework described in the methodology, we calculated composite performance scores for each LLM across all countries. Each dimension was assessed on a 0-5 rubric scale that evaluated quality and depth, with the complete rubric provided in the supplementary materials. The composite score formula: (Technical × 0.40) + (Contextual × 0.25) + (Skills × 0.20) + (Depth × 0.15) enabled standardized comparison across models and countries. The rubric-based contextual awareness score assesses how well responses integrated local context, complementing the binary indicator count in Section 5.2 that established whether specific elements were present at all.

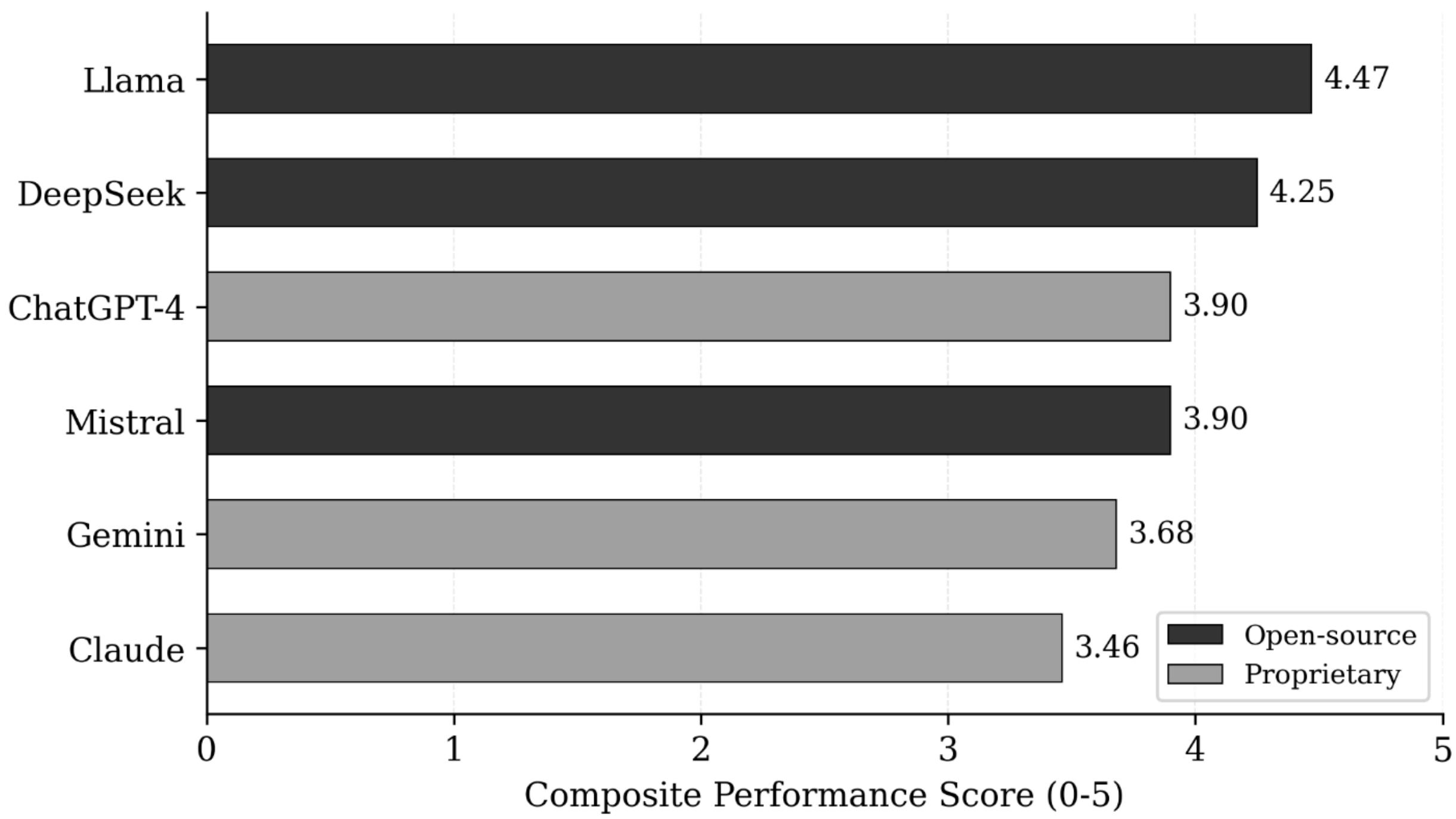

**Figure 4.** Composite performance scores for six LLMs across ten African countries. Scores calculated using weighted methodology: Technical Coverage (40%) + Contextual Awareness (25%) + Skills Balance (20%) + Depth (15%). Dark bars indicate open-source models; gray bars indicate proprietary models. Llama achieved the highest composite score (4.47/5) through our weighted methodology[1].

[1]for transparency, an example calculation: Llama's scores were Technical Coverage = 5.0, Contextual Awareness = 3.4 (average: 2,2,2,4,4,4,4,4,4,4 ÷ 10), Skills Balance = 5.0, Implementation Depth = 4.1. Applying weights: (5.0×0.40) + (3.4×0.25) + (5.0×0.20) + (4.1×0.15) = 4.47

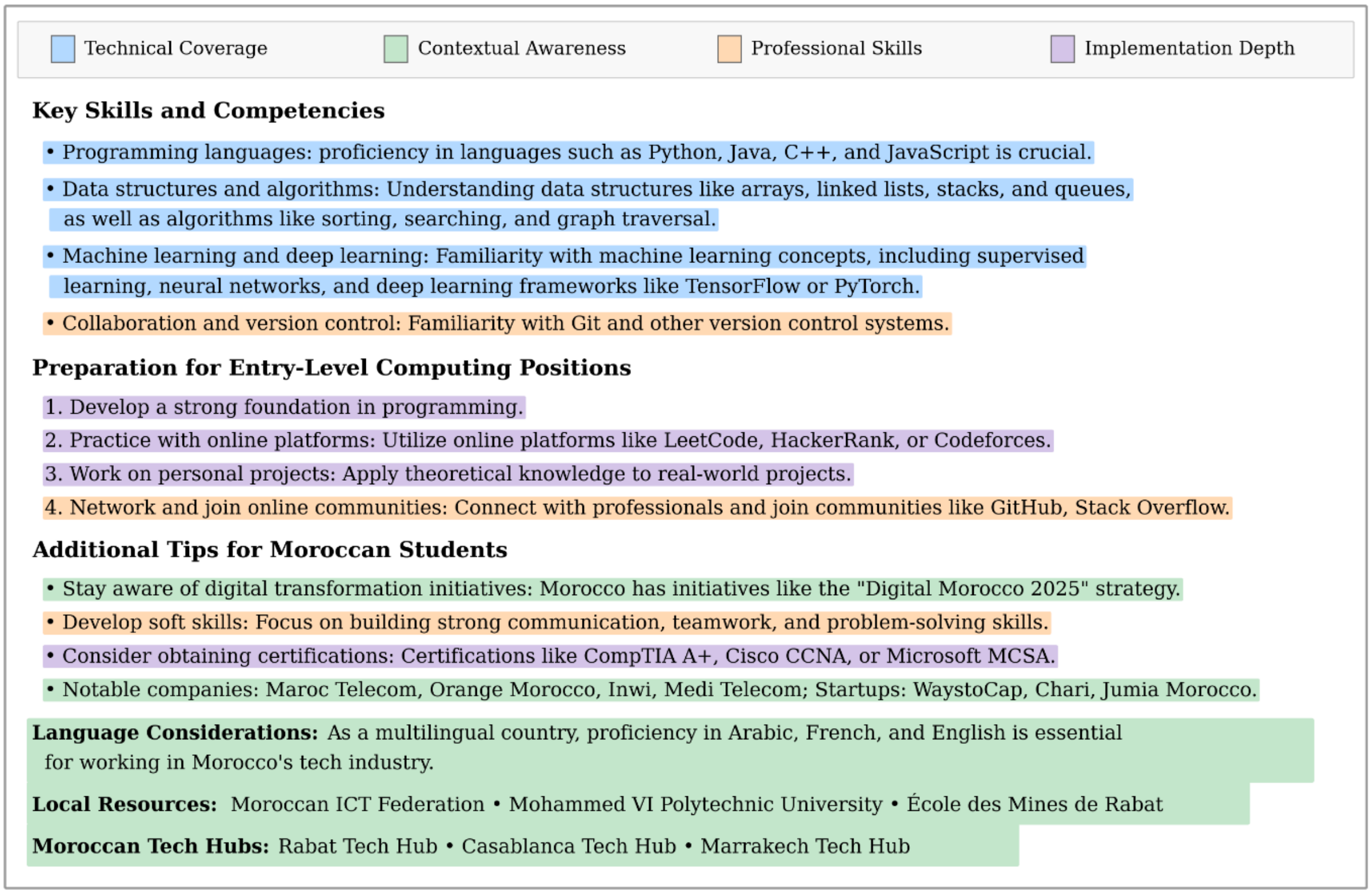

**Figure 5.** Excerpt from Llama response for Morocco illustrating evaluation dimensions and contextual references (composite score: 4.6/5).

Figure 5 illustrates the application of our weighted scoring methodology. The response achieved the highest overall performance score of 4.6/5 in our analysis. It earned perfect scores (5/5) in three dimensions: technical coverage demonstrated comprehensive competency identification across programming languages, AI/ML frameworks, cloud platforms, databases, cybersecurity, and NLP techniques; skills balance showed seamless integration of professional competencies like collaboration and version control alongside technical requirements; and implementation depth provided specific, actionable guidance including concrete platforms (LeetCode, HackerRank), certification pathways (CompTIA A+, Cisco CCNA), and detailed preparation strategies. However, contextual awareness received 4/5 rather than a perfect score because while the response included strong local elements like referencing “Digital Morocco 2025” strategy, specific companies like WaystoCap and Chari, multilingual requirements (Arabic, French, English), and regional tech hubs, it lacked references to local educational institutions or specific regulatory frameworks that would have warranted full points under our four-dimensional contextual framework. This evaluation approach, with its weighted calculation [$(5\times0.40) + (4\times0.25) + (5\times0.20) + (5\times0.15) = 4.6/5$], enabled objective comparison of how effectively different LLMs adapted their career guidance recommendations to diverse African educational and professional contexts across all ten countries examined.

Although Llama achieved the highest composite score compared to other models, DeepSeek emerged as the most successful model across countries, winning in 5 out of 10 countries (see Table 5 for country-specific rankings and margins). Notably, while Llama achieved the highest overall composite score through consistent performance across all dimensions, DeepSeek won in more individual countries (5 vs 4). This apparent contradiction reflects different performance patterns: Llama excelled through balanced consistency across all evaluation criteria, particularly in contextual awareness (3.4/5 vs DeepSeek's 2.2/5), while DeepSeek's country-specific victories were driven by exceptional scores in Implementation Depth (4.9/5) that proved particularly valuable in specific contexts. This illustrates how aggregate performance metrics may differ from context-specific effectiveness.

**Table 4. Country-Specific Model Performance Rankings**

| Country | Best Performing Model | Winner Score | 2nd Place Model | 2nd Score | Score Margin | Key Differentiator |
|---|---|---|---|---|---|---|
| **Tunisia** | Llama* | 4.75 | DeepSeek* | 4.75 | 0.00 | Perfect contextual awareness (4/4) with Digital Tunisia 2020 reference |
| **Ghana** | DeepSeek | 4.75 | Llama | 4.60 | 0.15 | Higher implementation depth (5.0 vs 4.0) |
| **Nigeria** | DeepSeek | 4.75 | Llama | 4.60 | 0.15 | Stronger technical coverage with local tech hub mentions |
| **Kenya** | Llama | 4.60 | DeepSeek | 4.25 | 0.35 | Higher contextual awareness (3/4 vs 2/4) – e.g., mentions of M-Pesa, local institutions |
| **Morocco** | Llama | 4.60 | ChatGPT-4 | 4.10 | 0.50 | Digital Morocco strategy reference, policy awareness (3/4) |
| **Benin** | Llama | 4.60 | DeepSeek | 4.25 | 0.35 | Higher contextual coverage, including mentions of "Digital Benin policy" |
| **Zambia** | Llama | 4.60 | DeepSeek | 4.25 | 0.35 | Stronger local institution recognition (3/4 vs 2/4) |
| **South Africa** | DeepSeek | 4.25 | Gemini/Mistral/Llama | 4.10 | 0.15 | Highest depth score (5.0) with multilingual context |

| Senegal | DeepSeek | 4.25 | Gemini/Mistral/Llama | 4.10 | 0.15 | Higher implementation guidance score (5.0 vs 4.0) |
|---|---|---|---|---|---|---|
| **Egypt** | ChatGPT-4 | 4.20 | Llama | 4.10 | 0.10 | Bilingual emphasis (Arabic/English), contextual score (2/4) |

*Tunisia shows a tie at 4.75 between Llama and DeepSeek, with Llama achieving higher contextual awareness (4/4 vs 3/4).

Llama topped the charts in 4 countries (Benin, Kenya, Morocco, Zambia) and tied with DeepSeek in Tunisia, while ChatGPT-4 achieved the best performance only in Egypt (4.2/5), due to its strong technical details. Mistral and ChatGPT-4 tied in composite scores (3.90/5), but Mistral achieved a perfect balance of skills across all countries while ChatGPT-4 showed stronger contextual adaptation.

Open-source models, Llama, DeepSeek, and Mistral outperformed their proprietary counterparts across both composite scores and the integration of skills. All three open-source models achieved full technical coverage (5.0/5), whereas proprietary models scored between 4.6 and 5.0. The most pronounced difference appeared in the skills balance dimension, where Llama and Mistral each received a perfect score (5.0/5), while proprietary models scored notably lower, ranging from 3.2 to 4.0.

## 6. Discussion

This discussion addresses our three RQs through analysis of our LLM-generated outputs. As an exploratory descriptive study, our findings identify patterns and generate hypotheses rather than establish causal relationships between model characteristics and output quality. The patterns we observed point to convergent technical recommendations alongside significant contextual gaps, raising broader questions about AI's role in African computing education.

### 6.1: Addressing RQ1 - Similarities in LLM Competency Descriptions

All LLMs agreed on core technical competencies, which suggests a key tension in global computing education between standardization and localization. Our findings demonstrated that Python programming and AI/ML/NLP appeared in all 60 responses (100% coverage) while cloud computing was mentioned in 47 responses (83% coverage), pointing to growing global agreement around certain skills. However, this apparent universality occurs within a complex postcolonial context where African computing education must balance the competing demands of international competitiveness with regional relevance (Tikly, 2019).

Although the focus on cloud platforms makes sense from a technical standpoint, it may unintentionally encourage dependence on foreign-built systems. Based on the outputs, proprietary models consistently assumed universal cloud access. For instance, ChatGPT-4's response for Nigeria stated that it would be valuable to: *"Gain experience with cloud services like AWS, Google Cloud, or Microsoft Azure, which are integral to deploying and scaling AI solutions."* Similarly, ChatGPT-4's response for Egypt advised students to *"Pursue relevant certifications, such as AWS Certified Solutions Architect or Google Associate Cloud Engineer,"* while Claude recommended for Morocco that students *"Obtain relevant certifications from cloud providers (AWS, Azure, GCP)."* These recommendations overlook documented barriers to certification access in many African contexts, including high examination costs, limited testing centers, and unequal access to reliable internet infrastructure (Blumberg et al., 2024). Studies on digital skills development have noted that local employers often prioritize practical experience and community-validated competencies over globally branded certifications alone (Bhorat et al., 2023; World Bank & IFC, 2025). Such patterns train students to use existing tools rather than develop their own, raising concerns about educational sovereignty.

The theoretical importance of these findings extends beyond listing or categorizing technical skills. It raises deeper questions about how knowledge is produced and how cultural epistemologies in computing education are represented. When LLMs repeatedly suggest the same technical competencies across dissimilar countries, they may reinforce what Santos (2007) describes as "epistemological monocultures" which represents dominant ways of thinking that exclude other valid approaches to learning and competency development. This uniformity appeared not only across models but also across countries, with competency recommendations showing little variation from one national context to another. This uniform approach can sideline local knowledge, such as designing for low-bandwidth environments or using indigenous languages, in favor of a single global idea of a competent graduate.

### 6.2: Addressing RQ2 - Differences in LLM Competency Descriptions and Model Performance Patterns

The higher scores of most open-source models challenge assumptions about AI tool quality, though performance varied considerably across individual models. As presented in our data (Fig. 3 & 4), open-source models generally outperformed proprietary ones across multiple dimensions, particularly in the balance between technical and professional skills, where the gap was most pronounced. However, this pattern was not uniform. Mistral, despite being open-source, scored 0.00/4 on contextual awareness, while ChatGPT-4 achieved 3/4 for Zambia. These exceptions caution against categorical claims about model types.

These findings have pedagogical implications for African education. The stronger contextual awareness of leading open-source models (Llama, DeepSeek) may reflect differences in training data and governance. Llama's response for Tunisia recognized the country's digital landscape, noting that *" Stay aware of the country's digital transformation initiatives: Tunisia has initiatives* *like the "Digital Tunisia 2020" strategy, which aims to drive economic growth and development through digital transformation."* Similarly, Llama's response for Kenya acknowledged the local fintech ecosystem: *"Understanding M-Pesa and mobile money technologies is valuable given Kenya's leadership in fintech innovation."* From a critical pedagogy perspective, open-source models may better embody democratic educational principles by providing accessible tools that can be adapted for local contexts rather than imposed as fixed solutions (Freire & Macedo, 2014). This finding is relevant for African educational institutions. It suggests that open-source tools, which are often more affordable, may offer not just cost savings but also better alignment with local needs. Accordingly, institutions might want to reconsider resource allocation, shifting funds from costly licenses toward infrastructure, training, and local innovation.

The noticeable variation in professional competency coverage (ranging from 27.5% to 97.5% across models) shows fundamental disagreements among AI systems about what constitutes comprehensive career preparation. Such variation resonates with ongoing debates in computing education about the relative importance of technical vs. professional, also known as "soft" or transferable competencies (Goleman, 2006; Gardner, 2006). From a holistic human development perspective, the finding that ethics and responsibility considerations appeared in only 43% of all responses is particularly concerning. In computing education, professional competencies including ethical awareness are fundamental to preparing graduates for responsible practice in increasingly complex technological environments (Clear et al., 2020). As African countries increasingly emphasize responsible AI development and ethical technology practices, computing graduates require a robust understanding of professional ethics, cultural sensitivity, and social responsibility. The inconsistent coverage of these competencies by LLMs suggests that current models cannot reliably support holistic professional development without deliberate interventions. In other words, if students were to rely solely on LLMs for guidance, it might paint an incomplete picture that underprepares them for teamwork, ethical dilemmas, or civic responsibilities in their careers.

The superior integration of professional competencies by open-source models may reflect different philosophical approaches to education embedded in their development. Open-source development communities often promote collaborative, community-oriented values (Raymond, 2001) that might translate into more well-rounded career guidance recommendations. Therefore, the governance structures and design philosophies of AI systems may significantly influence their outputs, a centrally important insight for policymakers deciding which tools to promote in academic settings. It could be beneficial for institutions to favor or customize AI tools that inherently value and include professional dispositions (for example, by fine-tuning models on data that contains not just technical questions and answers, but also content about teamwork, ethics case studies, etc.).

The notably poor contextual performance of Mistral (0.00/4 for contextual awareness) despite being open-source warrants closer examination. Mistral's smaller parameter size (7B) and efficiency focus may have prioritized technical accuracy over contextual adaptation. Unlike Llama (70B) and DeepSeek, which have substantially larger parameter counts and potentially more diverse training corpora, Mistral's design is optimized for computational efficiency rather than breadth of knowledge. As such, open-source development does not inherently guarantee contextual responsiveness; factors such as model scale, training data diversity, and design priorities may matter more than licensing approach alone. This exception underscores the importance of examining individual model characteristics rather than assuming categorical superiority based on development philosophy.

### 6.3: Addressing RQ3 - Comparative LLM Performance and Decolonizing Educational AI

The limited contextual awareness across most LLMs represents more than a technical limitation: it suggests that African technological contexts and educational settings are not well represented in the outputs of these AI tools. Our data showed that only 3 out of 60 responses achieved the maximum contextual awareness score (4/4), while 32 responses (53.3%) received zero scores for contextual elements. However, rapid advances in AI development mean these limitations may be addressed in future model versions, and our findings represent a snapshot of capabilities at one point in time. While our findings show an average contextual awareness of 35.4%, figuring out what counts as an 'acceptable' threshold requires more research involving African educators and students.

The limited recognition of local languages, cultural contexts, and indigenous knowledge in the majority of the LLM outputs suggests that these tools operate within what Mignolo (2011) describes as the "colonial matrix of power," which embodies a system that favors Western ways of knowing while sidelining others. The act of LLMs' generated responses overlooking the importance of multilingual skills for African computing professionals and local approaches to solving technical problems only serves to contribute to ongoing forms of educational colonialism, thereby making African educational systems increasingly reliant on Western technologies and content.

### 6.4 Contextual Blind Spots in LLM Career Guidance

Analysis of the raw LLM outputs revealed recurring contextual blind spots that reflect Western-centric assumptions about language, professional networking, and credentialing pathways. Should LLMs be expected to acknowledge local realities when providing career guidance? Given that guidance divorced from context is rarely actionable, contextual awareness is critical to the quality of recommendations (Trueblood et al., 2013). Digital Colonialism Theory clarifies why this matters: treating decontextualized advice as neutral obscures the assumptions embedded in mainstream training data. A human career counselor would naturally consider local hiring practices, language requirements, and available resources before advising a student (Fouad & Bingham, 1995). Contextually aware AI should meet the same standard. Three recurring blind spots illustrate this pattern.

**Linguistic Context and Multilingual Workplaces:** In Morocco, multiple LLM responses framed

language requirements in overly simplified terms. Gemini stated: "French and Arabic proficiency are essential in Morocco," while ChatGPT-4 advised: "As Morocco is bilingual (French and Arabic…), ensure you are comfortable communicating in these languages." However, Morocco's 2011 Constitution formally recognizes Amazigh (Berber) languages as official alongside Arabic (Kingdom of Morocco, 2011), and linguistic policy research documents persistent gaps between official recognition and implementation across education and public life (Ichou & Fathi, 2022). By framing Morocco as merely "bilingual," these responses obscure indigenous language realities and underrepresent multilingual competencies relevant in local workplaces.

**Professional Networking and Informal Hiring Practices:** Another recurring pattern concerned how models conceptualize professional networking. For Nigeria, Gemini recommended: "Online Platforms: Connect with professionals on LinkedIn and other platforms," while DeepSeek advised: "LinkedIn: Build a professional LinkedIn profile and connect with industry professionals." While formal networking platforms are widely used in Western labor markets, empirical research demonstrates that informal networks, personal referrals, and community contacts play a central role in job acquisition in many African economies, with informal job search channels frequently outperforming formal online platforms (World Bank, 2014). The exclusive focus on LinkedIn reflects a Western-centric model that may misrepresent how employment opportunities are actually accessed in local contexts.

**Infrastructure Assumptions and Technological Universalism:** Almost all models recommended cloud platforms requiring robust technological infrastructure, with an implicit assumption that African technological ecosystems are uniform and comparable to Western contexts. Many outputs assumed universal access to services like AWS or Azure while ignoring local laws and policies. This pattern echoes Graham's (2011) description of digital colonialism, wherein technology-based solutions are promoted without regard for local realities. Western-developed AI tools often carry built-in assumptions of high-resource settings, which, if adopted without critical evaluation, may widen the digital divide (Birhane, 2020; Kwet, 2019). By failing to consider infrastructure realities in many African contexts, LLM-generated advice can set expectations misaligned with conditions on the ground.

These excerpts illustrate how several LLMs rely on generalized, mainstream-oriented career advice templates even when prompted with country-specific contexts. Our comparative analysis revealed distinct patterns across model types. Proprietary models demonstrated what we term "contextual blindness," the consistent failure to acknowledge local realities despite country-specific prompting. This manifested as linguistic homogenization (assuming English-only contexts), technological universalism (recommending identical tools regardless of infrastructure), and regulatory ignorance (overlooking national policies and frameworks). Open-source models, while still limited, demonstrated greater contextual sensitivity, with Llama incorporating local company names, government initiatives, and technology ecosystems into recommendations. These patterns provide qualitative support for the quantitative finding of 35.4% average contextual awareness.

Llama's stronger contextual performance shows that more contextually responsive AI systems are technically possible. This suggests that the main barriers to improving contextual awareness are not technical limitations but rather the lack of diverse training data, design choices in algorithm development, and limited institutional commitment to building inclusive AI tools. Open-source models, which typically incorporate local data and expertise more readily, achieved better context alignment, implying that diversifying AI development teams and data sources through African-led AI initiatives could substantially close the contextual gap.

The gap between recommended technologies and on-the-ground realities carries significant implications for educational equity. Students who lack access to recommended tools and platforms because of financial constraints or infrastructure problems may face disadvantages in preparing for their careers, deepening existing inequalities. The concept of "appropriate technology" (Toboso et al., 2022) suggests that locally adapted solutions may be more sustainable and effective than importing high-resource solutions. From an educational perspective, this could mean focusing on teaching core principles using lightweight tools or offline resources, promoting open-source and offline-capable technologies, and encouraging ingenuity in low-resource settings as a strength rather than framing resource constraints as a deficit.

As African countries develop their own AI strategies and educational technologies, our findings suggest that locally developed or adapted AI tools may provide more contextually relevant educational guidance than globally standardized systems. This might mean investing in AI models trained or fine-tuned on region-specific data and involving educators from various African contexts in the development and evaluation process, aligning with decolonial paradigms that seek to create a more participatory innovation ecosystem.

Beyond these contextual blind spots, a broader tension emerged between the individualistic career advancement models promoted by most LLM responses and Ubuntu philosophy's emphasis on collective success and community responsibility (Letseka, 2013). This disconnect may leave African computing graduates navigating conflicting expectations between globally oriented technical training and locally valued professional conduct. In practice, students receiving LLM-generated guidance may be steered toward self-promotional strategies that feel misaligned with workplace cultures valuing relational trust and community contribution. Together with the patterns of epistemological reproduction and infrastructure assumptions discussed above, these findings suggest that without deliberate intervention, LLM-based career guidance risks reinforcing rather than disrupting existing inequities in global computing education.

### 6.5: Implications for African Computing Education Ecosystems

Collectively, our findings suggest the need for transformation in how African computing education institutions approach AI integration and career guidance, as well as the role of LLMs in students' preparation. Which tools are recommended and how they are used can define how well such suggestions may be actionable. Rather than adopting external AI tools as "neutral" technologies, institutions must critically evaluate these tools' contextual relevance, accessibility, and alignment with local educational values and industry needs.

### 6.5.1 Implications for Academic Institutions

The research points towards the value of hybrid approaches that combine the strengths of LLMs in identifying technical competencies with the contextual knowledge of local human experts. In practice, this means using LLMs to provide general insights on global industry trends, while relying on teachers, mentors, and community members to adapt that information to local needs. However, for this kind of integration to work well, it needs a solid understanding of both international computing standards and the cultural, linguistic, and institutional contexts where learning happens.

An Ubuntu-based perspective gives us a helpful way to rethink how AI gets integrated into African computing education. Based on the idea that technology should be developed with and for the community, Ubuntu encourages collaboration, shared responsibility, and empathy. So instead of relying only on external systems, institutions could involve students, teachers, and industry professionals in shaping AI tools together, making room for local voices and priorities (Omodan & Diko, 2021). This approach also changes the purpose of AI from just delivering answers to supporting relationships, like strengthening mentorship networks or building peer-to-peer learning, rather than the one-size-fits-all systems that deliver information from the top down.

For curriculum designers, these findings suggest the value of supplementing LLM-generated guidance with locally developed case studies and examples that highlight African technological innovations. Kenya's M-Pesa, for instance, transformed mobile money and financial inclusion across the continent (Ndung'u, 2021), while Nigerian startups have demonstrated African leadership in fintech infrastructure and global tech talent development. Computing programs might consider incorporating critical AI literacy modules that teach students to evaluate AI-generated advice against local realities, examining, for instance, whether recommended tools are accessible and whether suggested career pathways align with regional job markets. Career counselors and academic advisors should be aware that students are increasingly turning to LLMs for career guidance. Familiarity with both the capabilities and limitations of these tools is essential for providing informed support. This may require training programs that help counselors understand what LLMs typically recommend, where they fall short on contextual relevance, and how to guide students in critically evaluating AI-generated advice.

### 6.5.2 Implications for Industry and LLM Developers

For LLM developers and trainers, the findings point to the importance of diversifying training data to include African technology documentation, local policy frameworks, and region-specific career pathways. Fine-tuning approaches that incorporate feedback from African educators and industry professionals could help address the contextual gaps we identified, as could partnerships between global AI companies and African research institutions.

Furthermore, the better performance of open-source models suggests opportunities for African institutions to participate more actively in AI educational tool development rather than remain passive consumers of externally developed technologies. This participation could support broader goals of technological self-determination and educational sovereignty across the continent. If higher institutions of learning and local tech companies contribute to training or customizing LLMs through open collaborations or African-led AI coalitions, the resulting tools might inherently address the many gaps identified.

#### 6.5.3 Implications for Policy

The broader consequences of this analysis become evident in several areas requiring targeted policy attention.

**National Policy Alignment:** Based on the outputs, there is a critical misalignment between AI career guidance and specific African educational policies. South Africa's Protection of Personal Information Act (POPIA), fully enforceable since 2021, requires computing professionals to understand data protection compliance (Desai, 2024), yet only two models acknowledged such regulatory requirements. Kenya's National Digital Master Plan 2022-2032 and ICT in Education Policy 2021 emphasize digital literacy integration and local innovation ecosystems (TechMoran, 2022), which Llama's M-Pesa and Silicon Savannah references supported, but other models ignored. Nigeria's NITDA guidelines and Digital Economy Policy 2020-2030 prioritize fintech competencies and local technological capacity (NITDA, 2022), addressed only by Llama and DeepSeek.

**Strategic AI Adoption in African Higher Education:** African universities should prioritize open-source LLMs or, at minimum, critically assess AI-generated outputs rather than adopting proprietary tools based on their market reputation alone. The strong performance of cost-effective open-source models such as Llama (4.46/5) and DeepSeek (4.25/5) compared to proprietary alternatives like Claude (3.46/5) suggests resources would be better allocated toward improving infrastructure and developing locally relevant content. Institutions should adopt hybrid approaches that combine the technical strengths of LLMs in identifying global competencies with the contextual knowledge of local educators. This is essential for addressing the unrealistic infrastructure assumptions embedded in many AI tools, which often fail to reflect the conditions of African educational systems (Ade-Ibijola & Okonkwo, 2023).

**Industry Partnership Development:** The low contextual awareness scores highlight the need for stronger collaboration between universities and industry to improve the practical relevance of guidance tools. Partnerships can go beyond advisory boards to include co-designed training modules, joint curriculum reviews, and employer-led workshops that directly map to the skills most in demand in local job markets. Industry actors can also support graduate employability by offering more internships, hosting seminars, and creating hands-on opportunities where competencies are practiced in real settings. Ongoing feedback from these collaborations makes sure that AI tools are improved against actual workplace expectations rather than generic assumptions, producing guidance that is both accurate and locally meaningful for students across the continent.

**Pan-African Collaboration:** Continental cooperation, particularly efforts led by the African Union, can support the shared development of AI tools that reflect local cultures while staying flexible for national use. By pooling resources and expertise, it becomes easier for African countries to create systems that meet their own needs. This kind of cooperation also strengthens the continent's ability to shape its digital future and reduces the influence of foreign-based technologies that often carry biased or incomplete representations of African realities (Kwet, 2019; Mohamed et al., 2020).

**Ethical Implementation Requirements:** Educational technology must be designed and used with a clear understanding of the cultural and social settings where it operates. This means building local capacity for AI development, setting clear standards for cultural sensitivity, and actively involving users in shaping and evaluating the tools. These efforts help make sure that AI enhances rather than undermines African educational values, especially those grounded in community, care, and collective responsibility, as captured in the Ubuntu philosophy (Letseka, 2013).

## 7. Limitations and Future Directions

We acknowledge several limitations that should be considered when interpreting these findings, each of which also points toward opportunities for future research.

**Geographic Scope and Sampling**: Our country selection was based on higher rankings in AI readiness. This purposive choice enabled meaningful comparisons across more technologically advanced contexts but also introduced sampling bias, since countries with lower readiness levels, where issues of connectivity, affordability, and institutional support are often most acute, were not represented. These omissions are important given our critique of Western norms in AI tools. While this study focused on computing graduates in Africa, the methodological approach could be applied to other fields or regions from the Global South. Comparative studies might reveal similar patterns of digital colonialism in AI advice for Latin America or Southeast Asia, enriching global discourse on AI in education.

Additionally, Mauritius and Rwanda were omitted due to an oversight during the initial country mapping phase, identified only after data collection was complete. While both countries represent important technological contexts within Africa, incorporating them retrospectively was not feasible given significant LLM updates since data collection.

The focus on ten African countries and the computing discipline also limits the generalizability of findings to other countries or contexts. While the CC2020 framework offers a globally recognized standard for computing education, it may not fully reflect the realities of all African contexts or fields, requiring supplementation with locally developed competency frameworks for a more accurate and inclusive assessment. Going forward, we suggest expansion to additional countries and contexts to provide a more complete picture.

**Query Location and Model Behavior:** All responses were collected from Nigeria, providing an authentically African context but leaving open whether geolocation influenced outputs. Because some LLMs may adjust responses based on internet protocol (IP) address or routing, future research could explore variations using virtual private networks to test whether differences reflect model design or query location.

**Methodological Constraints:** The single standardized prompt approach, while necessary for systematic comparison, prevented models from demonstrating adaptive dialogue capabilities or contextual clarification that characterizes real-world LLM usage. This design choice also meant that variations in prompt phrasing, which may influence outputs, could not be examined. Future studies should employ inferential statistics with multiple queries per condition and larger samples to test whether the performance differences observed are statistically significant.

**Language and Cultural Context:** The English-only analysis represents a significant limitation given Africa's multilingual professional context, where computing careers frequently require proficiency in indigenous languages, French, Portuguese, or Arabic alongside English. By restricting analysis to English, the study cannot assess how LLMs perform when prompted in other languages common across the continent, nor can it evaluate whether models provide culturally appropriate guidance for non-Anglophone career contexts. This constraint likely affects the generalizability of our contextual awareness findings. Exploring LLM responses in African languages, including code-switching scenarios common in African workplaces, could illuminate how language and culture influence AI effectiveness.

**Analytical Approach: T**his exploratory study does not include inferential statistical testing, validation with end-users, or comparison with human counselors. Findings should be interpreted as descriptive patterns warranting confirmatory research rather than definitive conclusions about model superiority. Longitudinal studies comparing students using AI guidance with those using traditional counseling could test the impact of LLM-generated advice on actual graduate outcomes.

**Temporal Validity:** LLM capabilities evolve rapidly through continuous training updates and architectural improvements. Our findings represent a snapshot from April 2025 and should be interpreted as capturing model capabilities at that specific point. Given the approximately 3–6-month update cycle for major LLMs, readers should consider that model capabilities may have evolved by publication time. However, the structural patterns identified, particularly regarding infrastructure assumptions and contextual gaps, likely reflect deeper training data limitations that may persist across model versions.

**Interpretive Scope:** Our interpretation of contextual relevance is shaped by our understanding of African educational contexts, which may not capture the continent's full diversity. Despite achieving acceptable inter-rater reliability, qualitative interpretation inherently involves subjective judgment. There is also a layer of complexity in African researchers studying AI systems created primarily in Western countries. While combining Western academic methods with African philosophical values can enrich analysis, it may also create tensions, as the two approaches rest on different worldviews. Future studies should incorporate direct input from African students, educators, and employers to validate whether the contextual gaps we identified align with stakeholder perceptions and priorities.

**Subgroup and Equity Considerations:** This study did not examine how LLM career guidance might differ based on user characteristics such as gender, socioeconomic background, or level of digital access. Career pathways and barriers differ substantially across these dimensions in African contexts, and guidance that emphasizes expensive certifications or assumes reliable internet access may be less appropriate for students from lower-income backgrounds or those in regions with limited connectivity. Future studies should explore whether LLM outputs vary when prompts specify user demographics and whether certain recommendations systematically disadvantage particular groups.

**Infrastructure Assumptions:** Although this study identified pervasive infrastructure assumptions, it did not include empirical testing in real-world African educational settings. The gap analysis was based on existing literature, rather than direct fieldwork across different connectivity levels. Research should investigate AI tools designed for low-resource settings, such as LLMs that run locally on mobile phones or provide differentiated responses based on connectivity limitations.

Broader Research Opportunities: Beyond addressing these specific limitations, several broader research opportunities emerge. Developing a standardized framework for evaluating contextual responsiveness would ensure consistency across future studies in diverse African contexts. Experimental field trials could determine optimal sequencing of AI and human input in hybrid career counseling approaches. Research integrating Ubuntu-based pedagogies or indigenous African knowledge into AI training data could advance understanding of how AI might embody non-Western philosophies (Chilisa, 2017). Investigating how students perceive AI advice, including their trust levels and effects on self-efficacy and career aspirations, would also be valuable. If AI consistently undervalues professional competencies or local context, this may subtly discourage students from pursuing those areas. Understanding user psychology will be important for developing both AI systems and critical AI literacy among users

These limitations collectively mean that our findings should be interpreted as initial exploratory insights rather than definitive evidence of model capabilities. The patterns we identify provide hypotheses for future research rather than conclusive judgments about LLM suitability for African educational contexts. We encourage readers and practitioners to consider the findings in light of these constraints. We see this research as a starting point toward equitable, contextually relevant AI in education. Steering AI adoption toward innovations that empower rather than marginalize will require continued inquiry rooted in both global knowledge and local wisdom.

## 8. Conclusion

Our study provided the first comprehensive evaluation of LLMs as career guidance tools for computing students across ten African countries. The most significant finding was the contextual awareness gap: only 35.4% average contextual relevance, with over half of responses receiving zero scores for local context. While LLMs reliably identified universal technical competencies, they consistently failed to address African-specific educational policies, infrastructure realities, and professional norms. These patterns reflected the Digital Colonialism dynamics identified in our framework, where Western-centric training data shaped AI outputs that inadequately served African contexts. Notably, open-source models outperformed proprietary alternatives, though Mistral's poor contextual performance, despite being open-source, demonstrated that open development alone does not guarantee cultural responsiveness. These findings suggested that African educational institutions should critically evaluate AI tools rather than adopting them based on market reputation, and that hybrid approaches combining AI technical strengths with local human expertise merit further development. As AI continues to reshape educational practice, ensuring that these tools effectively support African computing students in their career preparation will require sustained attention to contextual responsiveness, participatory development, and critical evaluation frameworks grounded in local realities.

## Data Availability Statement

The codebook, coded dataset, and supporting materials for this study are available on Mendeley Data: https://data.mendeley.com/datasets/xt42c2jfdg/1

**APPENDIX A**: Example Prompts and Representative LLM Responses

**Standardized Prompt:**

"What large language model skills or competencies do students applying for computing jobs in [Country] need to know and how should they prepare for entry-level computing positions?

**Example 1: Llama Response for Tunisia (Contextual Awareness Score: 4/4)**

Introduction to Large Language Model Skills
As a Meta AI assistant, I can provide insights on the essential skills and competencies required for students applying for computing jobs, particularly in Tunisia. To increase their chances of securing entry-level computing positions, students should focus on developing the following large language model skills:
Key Skills and Competencies

- Programming languages: Proficiency in languages such as Python, Java, C++, and JavaScript is crucial.
- Data structures and algorithms: Understanding data structures like arrays, linked lists, stacks, and queues, as well as algorithms like sorting, searching, and graph traversal.
- Machine learning and deep learning: Familiarity with machine learning concepts, including supervised and unsupervised learning, neural networks, and deep learning frameworks like TensorFlow or PyTorch.
- Natural Language Processing (NLP): Knowledge of NLP techniques, including text preprocessing, sentiment analysis, and language modeling.
- Cloud computing: Understanding of cloud platforms like AWS, Azure, or Google Cloud, including their services and applications.
- Database management: Familiarity with database concepts, including data modeling, normalization, and querying using SQL or NoSQL databases.
- Cybersecurity: Basic understanding of security principles, including authentication, authorization, and encryption.
- Collaboration and version control: Familiarity with Git and other version control systems, as well as collaboration tools like GitHub or Bitbucket.

Preparation for Entry-Level Computing Positions
To prepare for entry-level computing positions in Tunisia, students should:

1. Develop a strong foundation in programming: Focus on building a solid understanding of programming concepts, data structures, and algorithms.
2. Practice with online platforms: Utilize online platforms like LeetCode, HackerRank, or Codeforces to practice coding challenges and improve problem-solving skills.
3. Work on personal projects: Apply theoretical knowledge to real-world projects, demonstrating skills and experience to potential employers.
4. Stay updated with industry trends: Follow industry leaders, researchers, and blogs to stay informed about the latest developments and advancements in the field.
5. Pursue internships or certifications: Consider internships or certifications to gain practical experience and demonstrate expertise to potential employers.
6. Network and join online communities: Connect with professionals and join online communities like GitHub, Stack Overflow, or Reddit to learn from others and stay informed about job opportunities.

Additional Tips for Tunisian Students

- Familiarize yourself with the Tunisian job market: Research the top companies, job roles, and required skills in Tunisia's computing industry.
- Learn about local industry trends: Stay updated on the latest developments and advancements in Tunisia's tech industry, including emerging technologies and innovations.
- Develop soft skills: Focus on building strong communication, teamwork, and problem-solving skills, which are essential for success in the computing industry.
- Consider obtaining certifications: Certifications like CompTIA A+, Cisco CCNA, or Microsoft MCSA can be beneficial in increasing job prospects.
- Stay aware of the country's digital transformation initiatives: Tunisia has initiatives like the "Digital Tunisia 2020" strategy, which aims to drive economic growth and development through digital transformation.

Some notable companies in Tunisia's tech industry include:

- Tunisie Telecom
- Orange Tunisia
- Ooredoo Tunisia
- Société Tunisienne de l'Electricité et du Gaz (STEG)
- Tunisian Startups, such as Expensya, InstaDeep, and Kaoun

By focusing on these skills and competencies, students in Tunisia can increase their chances of securing entry-level computing positions and launching a successful career in the field.

Language Considerations

As a multilingual country, proficiency in Arabic, French, and English is essential for working in Tunisia's tech industry. Students should aim to be proficient in at least two of these languages to increase their job prospects.

Local Resources

- Tunisian Startup Ecosystem: A platform that connects startups, investors, and mentors.
- Tunisian ICT Chamber: A professional organization that promotes the development of the ICT sector in Tunisia.
- École Polytechnique de Tunisie: A leading institution for engineering and technology education in Tunisia.

By leveraging these resources and focusing on the key skills and competencies, students in Tunisia can successfully prepare for entry-level computing positions and contribute to the country's growing tech industry.

**Example 2: Mistral Response for Nigeria (Contextual Awareness Score: 0/4)**

Students applying for computing jobs in Nigeria, particularly those involving large language models (LLMs), need to develop a specific set of skills and competencies to prepare for entry-level positions. Here are some key areas to focus on:

1. Natural Language Processing (NLP): Understanding NLP is crucial as it enables computers to interpret and generate human language. This includes tasks like sentiment analysis, named entity recognition, text classification, and machine translation 13.
2. Machine Learning and Deep Learning: Familiarity with machine learning algorithms and deep learning techniques is essential. This includes knowledge of supervised and unsupervised learning, neural networks, and specific models like recurrent neural networks (RNNs) and transformers 13.
3. Programming Skills: Proficiency in programming languages such as Python is important, as it is widely used in AI and NLP development. Knowledge of machine learning frameworks like TensorFlow or PyTorch is also beneficial 13.
4. Cloud Computing: Many computing jobs require knowledge of cloud computing platforms such as

AWS, Azure, or Google Cloud. Experience in designing, developing, and managing cloud-based infrastructure and applications is valuable 14.
5. Soft Skills: Strong analytical and problem-solving abilities, excellent communication, and collaboration skills are essential for working effectively in a team and contributing to projects 13.
To prepare for these roles, students should:
• Gain Practical Experience: Work on NLP projects, participate in NLP challenges, and contribute to open-source projects to build a strong portfolio.
• Stay Updated: Join online communities, attend meetups and conferences, and follow the latest advancements in the field.
• Certifications: Consider obtaining certifications in cloud computing or relevant AI technologies to enhance your credentials.
By focusing on these skills and preparation strategies, students can better position themselves for entry-level computing jobs in Nigeria.

**Example 3: ChatGPT-4 Response for Egypt (Contextual Awareness Score: 2/4)**

To excel in Egypt's evolving computing job market, especially in roles involving large language models (LLMs) and artificial intelligence (AI), students should focus on developing the following key skills:
1. Proficiency in Artificial Intelligence and Machine Learning:
• Understanding AI Fundamentals: Grasp the basics of AI, including machine learning algorithms, natural language processing, and model evaluation techniques.
• Practical Application: Engage in projects that involve training and deploying AI models to solve real-world problems.
2. Cloud Computing Expertise:
• Familiarity with Cloud Platforms: Gain experience with cloud services like AWS, Google Cloud, or Microsoft Azure, which are integral to deploying and scaling AI solutions.
• Certifications: Pursue relevant certifications, such as AWS Certified Solutions Architect or Google Associate Cloud Engineer, to validate your skills.
3. Cybersecurity Knowledge:
• Understanding Security Protocols: Learn about data protection, encryption, and secure coding practices to safeguard AI applications.
• Staying Updated: Keep abreast of the latest cybersecurity threats and mitigation strategies.
4. Software Development Skills:
• Programming Languages: Achieve proficiency in languages commonly used in AI development, such as Python, Java, or C++.
• Software Engineering Principles: Understand version control, testing, and agile methodologies to contribute effectively to development teams.
5. Language Proficiency:
• English Proficiency: Since English is the primary language in the tech industry, strong reading, writing, and speaking skills are essential.
• Arabic Fluency: In Egypt, being fluent in Arabic can provide a significant advantage, as many employers value bilingual candidates.
Nucamp
Preparation Strategies:
• Enroll in Specialized Training Programs: Participate in initiatives like PwC's "Jump Start Your Career" program, which offers certifications in AWS and SAP, along with training in business and consulting skills.
PwC
• Pursue Relevant Certifications: Obtaining certifications in AI, cloud computing, and cybersecurity can enhance your employability and demonstrate your commitment to the field.

- Engage in Continuous Learning: Stay updated with the latest industry trends by attending workshops, webinars, and conferences related to AI and computing.
- Develop a Strong Portfolio: Work on projects that showcase your skills in AI and software development, and consider contributing to open-source projects to gain practical experience.
- Enhance Language Skills: Invest time in improving both English and Arabic proficiency to communicate effectively in professional settings.

By focusing on these competencies and preparation strategies, students can position themselves competitively for entry-level computing positions in Egypt's dynamic tech landscape.

## Appendix B: Contextual Indicator Presence by Country and Model

Table 3b: Contextual Indicator Presence by Country and Model

| Country | Model | Tech Industry | Language/ Culture | National Policy | Local Institution | Score |
|---|---|---|---|---|---|---|
| Egypt | ChatGPT-4 | 1 | 1 | 0 | 0 | 2 |
| | Claude | 0 | 0 | 0 | 0 | 0 |
| | Gemini | 0 | 0 | 0 | 0 | 0 |
| | Mistral | 0 | 0 | 0 | 0 | 0 |
| | Llama | 1 | 0 | 0 | 0 | 1 |
| | DeepSeek | 0 | 0 | 0 | 0 | 0 |
| South Africa | ChatGPT-4 | 1 | 0 | 0 | 0 | 1 |
| | Claude | 1 | 1 | 0 | 0 | 2 |
| | Gemini | 1 | 0 | 0 | 0 | 1 |
| | Mistral | 0 | 0 | 0 | 0 | 0 |
| | Llama | 1 | 1 | 0 | 0 | 2 |
| | DeepSeek | 1 | 1 | 0 | 0 | 2 |
| Senegal | ChatGPT-4 | 1 | 0 | 0 | 0 | 1 |
| | Claude | 1 | 1 | 0 | 0 | 2 |
| | Gemini | 1 | 1 | 0 | 0 | 2 |
| | Mistral | 0 | 0 | 0 | 0 | 0 |
| | Llama | 1 | 1 | 0 | 0 | 2 |
| | DeepSeek | 1 | 1 | 0 | 0 | 2 |
| Tunisia | ChatGPT-4 | 1 | 0 | 0 | 0 | 1 |
| | Claude | 1 | 1 | 0 | 0 | 2 |
| | Gemini | 1 | 1 | 0 | 0 | 2 |
| | Mistral | 0 | 0 | 0 | 0 | 0 |
| | Llama | 1 | 1 | 1 | 1 | 4 |
| | DeepSeek | 1 | 1 | 0 | 1 | 3 |
| Kenya | ChatGPT-4 | 1 | 0 | 0 | 0 | 1 |
| | Claude | 1 | 1 | 0 | 0 | 2 |
| | Gemini | 1 | 0 | 0 | 0 | 1 |
| | Mistral | 0 | 0 | 0 | 0 | 0 |
| | Llama | 1 | 1 | 0 | 1 | 3 |
| | DeepSeek | 1 | 1 | 0 | 0 | 2 |
| Nigeria | ChatGPT-4 | 1 | 0 | 0 | 0 | 1 |
| | Claude | 1 | 1 | 0 | 0 | 2 |
| | Gemini | 1 | 0 | 0 | 0 | 1 |
| | Mistral | 0 | 0 | 0 | 0 | 0 |
| | Llama | 1 | 1 | 0 | 1 | 3 |
| | DeepSeek | 1 | 1 | 0 | 1 | 3 |
| Ghana | ChatGPT-4 | 1 | 0 | 0 | 0 | 1 |
| | Claude | 1 | 1 | 0 | 0 | 2 |
| | Gemini | 1 | 0 | 0 | 0 | 1 |
| | Mistral | 0 | 0 | 0 | 0 | 0 |
| | Llama | 1 | 1 | 1 | 1 | 4 |
| | DeepSeek | 1 | 1 | 0 | 1 | 3 |
| Benin | ChatGPT-4 | 1 | 1 | 0 | 0 | 2 |
| | Claude | 0 | 0 | 0 | 0 | 0 |
| | Gemini | 0 | 0 | 0 | 0 | 0 |
| | Mistral | 0 | 0 | 0 | 0 | 0 |
| | Llama | 1 | 1 | 1 | 0 | 3 |
| | DeepSeek | 1 | 1 | 0 | 0 | 2 |
| Zambia | ChatGPT-4 | 1 | 0 | 0 | 0 | 1 |
| | Claude | 1 | 1 | 0 | 0 | 2 |
| | Gemini | 0 | 0 | 0 | 0 | 0 |
| | Mistral | 0 | 0 | 0 | 0 | 0 |
| | Llama | 1 | 1 | 0 | 1 | 3 |
| | DeepSeek | 1 | 1 | 0 | 0 | 2 |
| Morocco | ChatGPT-4 | 1 | 1 | 0 | 0 | 2 |
| | Claude | 1 | 1 | 0 | 0 | 2 |
| | Gemini | 1 | 0 | 0 | 0 | 1 |
| | Mistral | 0 | 0 | 0 | 0 | 0 |
| | Llama | 1 | 1 | 1 | 0 | 3 |
| | DeepSeek | 1 | 0 | 0 | 0 | 1 |

*Note: 1 = indicator present; 0 = indicator absent. Score = sum of indicators present (maximum = 4).*